\documentclass[5p,twocolumn,10pt]{elsarticle}

\usepackage{lineno,hyperref}
\modulolinenumbers[5]

\journal{Journal of Process Control}

\usepackage{amsmath} 			% equation, align, gather, flalign, multline, alignat, split
\usepackage{amssymb} 			% symbols
\usepackage{cases} 				% numcases
\usepackage{siunitx}			%korrektes Darstellen von Einheiten
\usepackage{cases} 				% numcases
\usepackage{siunitx}			%korrektes Darstellen von Einheiten
\usepackage{etoolbox}
\usepackage{kbordermatrix}
\usepackage{enumitem}
\usepackage{calc}
\usepackage{tikz}
\usetikzlibrary{shapes}

\newcommand*{\pd}[3]{\left(\frac{\partial\, #1}{\partial\, #2}\right)_{\!\!\!#3}}
\newcommand*{\diff}[1]{\text{d}\, #1}
\newcommand*{\difffrac}[2]{\frac{\diff{#1}}{\diff{#2}}}
\newcommand*{\fcn}[2]{#1 \left(\,#2\,\right)}
\newcommand*{\fcnII}[2]{#1 \{\,#2\,\}}
\newcommand*{\smallfcn}[2]{#1\! \left(#2\right)}

\newcommand*{\md}{\dot{m}}

\newcommand*{\Tr}{{\bar{T}}}
\newcommand*{\pr}{{\bar{p}}}
\newcommand*{\vr}{{\bar{v}}}

\newcommand*{\ur}{{\bar{u}}}
\newcommand*{\hr}{{\bar{h}}}
\newcommand*{\cvr}{{\bar{c_v}}}

\newcommand*{\Mu}{\text{M}_{d_2}}
\newcommand*{\Dmu}{\Delta\mu}
\newcommand*{\Dphi}{\Delta\varphi}
	
\newcommand*{\Pa}[1]{#1^{\text{P0}}}	
\newcommand*{\Pb}[1]{#1^{\text{P1}}}	
\newcommand*{\Pc}[1]{#1^{\text{P2}}}	
\newcommand*{\Pj}[1]{#1^{\text{Pj}}}
\newcommand*{\Pjm}[1]{#1^{\text{P[j-1]}}}
\newcommand*{\PN}[1]{#1^{\text{PN}}}	
\newcommand*{\Sa}[1]{#1^{\text{S1}}}	
\newcommand*{\Sb}[1]{#1^{\text{S2}}}	
\newcommand*{\Sj}[1]{#1^{\text{Sj}}}	
\newcommand*{\Si}[1]{#1^{\text{Si}}}
\newcommand*{\Sjm}[1]{#1^{\text{S[j-1]}}}
\newcommand*{\Sjp}[1]{#1^{\text{S[j+1]}}}
\newcommand*{\SN}[1]{#1^{\text{SN}}}	

\newcommand{\omomR}{\frac{\omega}{\omega_R}}
\newcommand*{\tipspeed}{\left[\pi\,n\, d_2\right]}	
	
\renewcommand*{\vec}[1]{\underline{#1}}
\newcommand*{\zk}[1]{\vec{#1}_k}	
\newcommand*{\zkm}[1]{\vec{#1}_{k-1}}	
\newcommand*{\xk}{\zk{x}}	
\newcommand*{\xkm}{\zkm{x}}	
\newcommand*{\mat}[1]{\mathbf{#1}}

\newcommand*{\Az}{\mat{A}_k}
\newcommand*{\bz}[1]{\vec{b}_k^{{#1}}}
\newcommand*{\bv}{\bz{\varphi}}
\newcommand*{\bm}{\bz{\mu}}

\newcommand*{\PjTv}[2]{(\Pj{\Tr_{#1}},\Pj{\vr_{#2}})}

\newcommand{\negphantom}[1]{\settowidth{\dimen0}{$#1$}\hspace*{-\dimen0}}

\newlength{\saveleftmarginA}
\newlength{\saveleftmarginB}
\newlength{\saveleftmargin}

\newcommand*{\COO}{CO\textsubscript{2}}

\allowdisplaybreaks

\begin{document}

\begin{frontmatter}
	
	\title{DRAFT: Real-Time Estimation of a Multi-Stage Centrifugal Compressor Performance Map Considering Real-Gas Processes and Flexible Operation}
	
	\author{Maik Gentsch\corref{cor1}%
		\fnref{fn1}}
	\ead{maik.gentsch@tu-berlin.de}
	
	\author{Rudibert King\fnref{fn2}}
	\ead{rudibert.king@tu-berlin.de}
	
	\cortext[cor1]{Corresponding author}
	\fntext[fn1]{Graduate Research Assistant; Declarations of interest: none}
	\fntext[fn2]{Head of Department; Declarations of interest: none \\ \textcopyright\ 2019. This manuscript version is made available under the CC BY-NC-SA 4.0 license \url{https://creativecommons.org/licenses/by-nc-sa/4.0/}\ .}
%	\fntext{}
	
	\address{Technische Universit\"at Berlin, Department of Measurement and Control, Hardenbergstr.\ 36a, 10623 Berlin, Germany}


	\begin{abstract}
		This paper contributes to modeling and supervision of multi-stage centrifugal compressors coping with real-gas processes and steady to highly transient operating conditions.
		A novel dynamic model is derived, and the incorporation of the generic \textsc{Lee-Kesler-Pl\"ocker} real-gas equation of state and its derivatives is presented.
		The model allows for embedding arbitrarily shaped performance maps, based on state-of-the-art polytropic change-of-state compressor characteristics.
		As the validity of these maps is a key issue for simulation and model-based monitoring, performance maps are treated as time-variant, and their shape is to be identified and monitored during operation.
		The proposed real-time map estimation scheme comprises an Unscented Kalman Filter and a newly proposed algorithm, referred to as Recursive Map Estimation.
		The combination yields a novel parameter and state estimator, which is expected to be superior if some parameters are characterized by a distinct operating point dependency. 
		Two additional time-variant parameters are provided for monitoring: The first indicates the level of confidence in the local estimate, and the second points to drastic performance map alterations, which may be further exploited in fault detection.
		A modified reference simulation of a two-stage supercritical carbon dioxide compressor with known state trajectories, performance maps, and alterations demonstrates the successful application of the entire monitoring scheme, and serves for a discussion of the results.
	\end{abstract}
	

\end{frontmatter}

%%%%%%%%%%%%%%%%%%%%%%%%%%%%%%%%%%%%%%%%%%%%%%%%%%%%%%%%%%%%%%%%%%%%%%
\section{Introduction}
\label{sec:intro}

% Request of flexibility
Off-design, and in particular, flexible operation of industrial plants arises as a consequence of economic interests, and the integration of volatile, highly dynamic fossil-free resources in power generation that has to be complemented for by quickly responding conventional gas turbines.
%The latter affects several plants or plant components due to the volatile nature of regenerative power production as conventional power production by means of gas turbines has to step in quickly or 
Likewise, industries start to dynamically adapt their production to the current prizes of the energy market, e.g., in air separation.
This again results in the dynamic operation of the air compressors used.
Consequently, the design philosophy, as well as the supervision and maintenance of these machines, has to be adapted properly.
Concerning supervision for flexible operation, the common ``steady-state'' assumption is likely to provide erroneous results, e.g., frequent false alarms, if the plant's transient behavior comprises any dynamic within a relevant time scale.
The same issue applies if inappropriate model assumptions are used, e.g., the integration of the ideal-gas equation for real-gas processes in supervision algorithms.

\begin{figure*}
	\centering
	\includegraphics{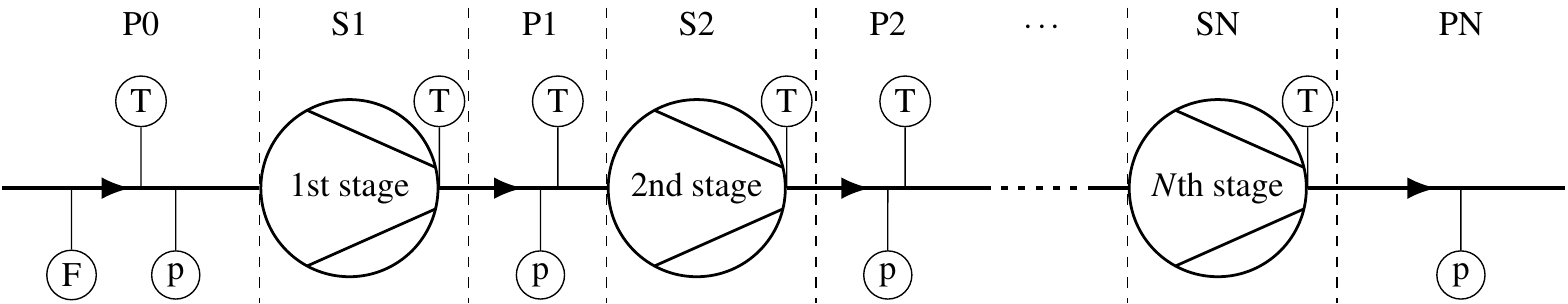}
	\caption[Compression unit]{Compression unit with mass flow (F), temperature (T), and pressure (p) instrumentation}
	\label{fig:unit}
\end{figure*}

In this context, this paper deals with modeling (see Section \ref{sec:model}) and supervision (see Section \ref{sec:monitoring}) of multi-stage (centrifugal) compressors coping with real-gas processes and flexible operation.
For geared compressors, these machines comprise large pipes as the connection between (compressor stage)--(compressor stage), (compressor stage)--(intercooler), (compressor stage)--(valve), etc.
The strategy pursued here is picking out (multi-stage) compression units that are not interrupted by other plant components (intercooler, valves, etc.).
For such compression units (Fig.\ \ref{fig:unit}), a model structure is introduced in Section \ref{sec:structure}.
The other plant components could be included readily in the concept proposed, but this is not done here.
Likewise, to keep the presented equations compact, additional dependence on potentially installed inlet guide vanes is discarded.
%However, if the basic conceptual idea is conveyed, merely some diligent effort is necessary to include such a dependency.

The proposed model will be applied within the model-based monitoring scheme.
But it might be used for dynamic process simulations as well, as it is designed to cope with highly flexible operating conditions.
However, for the simulation task, a quasi-steady-state behavior of a compressor stage is assumed, i.e., the validity of a compressor stage specific performance map remains unaffected, even for transients.
The investigation in \cite{blieske2011centrifugal} supports this general practice, with the exception for power calculated from such a static map.
A multitude of publications deals with the concrete shape of such maps, i.e., the concrete correlations between head, compression work, volume flow, and speed, or respective representatives of this compressor characteristics.
In \cite{casey2012method}, a set of equations is presented that aims to describe the performance map far away from the design point.
As in the present contribution, the methodology in \cite{casey2012method} is based on a static dependence of the same dimensionless compressor characteristics as they are used here.
%Truly, there is a great demand for predicting a map in uncertain regions that are not accessed in experiments or far away from the certainly known behavior near the design point.
Similarly, a map prediction and modification scheme is proposed in \cite{bayomi2013centrifugal}.
The prediction is based on several models that are originally introduced by \textsc{Moore} and \textsc{Greitzer} \cite{moore1986theory}, \textsc{Dixon} \cite{dixon2013fluid}, and \textsc{Gravdahl} \cite{gravdahl1998modeling}.
An exhaustive description of traditional models for axial and centrifugal compressors can be found in \cite{gravdahl2012compressor}.
However, starting point for a simulation with the dynamic model proposed herein, is a given set of discrete operating points, which may be derived from the mentioned approaches.
The current operating point within the performance map is then calculated via an interpolation scheme.
This enables the integration of arbitrarily shaped maps, facilitating a high degree of freedom for the task of learning a map in the framework of the model-based monitoring proposed.

%The presented monitoring routine is capable of 

As mentioned in \cite{ludtke2004process}, almost all process fluids that are used in centrifugal compressors have distinct real-gas behavior.
This applies in particular to carbon dioxide (\COO), and to an even greater extent, to supercritical \COO, which has to be handled in a \textit{Carbon Capture and Storage} application, for example.
With the purpose of providing a versatile compressor model, the generic and easy-to-parametrize \textsc{Lee-Kesler-Pl\"ocker} (LKP) real-gas model (see \cite{lee1975LKP}) is embedded in this contribution.
% whereas the case study considered will actually focus on supercritical \COO.
The LKP model is capable of describing the thermal relations for a multitude of relevant process fluids and conceivable mixtures properly.
The specific integration of the respective equation of state and its derivatives is introduced in Section \ref{sec:realgas}.
%For the validation of the presented model-based monitoring scheme (s. Section \ref{sec:results}), the supercritical \COO\ case is selected.

Subject to the existence of the instrumentation, depicted in Fig.\ \ref{fig:unit}, the monitoring scheme developed in this contribution is capable of tracking common compressor characteristics (polytropic head, efficiency, etc.) separately for each compressor stage.
Moreover, several fluid temperature estimates are provided, which is desired for considerably delayed temperature measurements, as is the case for many high-pressure applications.
Since the calculation of compressor characteristics is based on stage-specific performance maps (see Section \ref{sec:stagemap}), which can change, e.g., due to fouling, or which are not known exactly initially, the validity of these maps is a key issue for simulation and model-based monitoring.
Therefore, performance maps are treated as time-variant, and the proposed approach aims to identify and monitor their shapes during operation.
Online adaption of performance maps for centrifugal compressors has been presented in \cite{cortinovis2014online}.
There, an automated decision unit, based on the deviation between measurements and the current map, triggers a (sequential) quadratic program from time to time to calculate the new map.
Therefore, a proper set of past measurements is stored in a buffer.
In contrast to this batch approach, which is not stated to be designed for real-gas processes and transient operation, the proposed algorithm in this contribution is characterized by a constant computational load and storage requirement for every time interval between two measurement samples.
Likewise, this applies to the work of \textsc{H\"ockerdal} et al.\  \cite{hoeckerdal2011ekf}.
The core concept of their contribution, as is the case for the monitoring scheme here, is the preservation of the operating point dependence of parameters via estimated grid points of a parameter map.
Their approach is a joint estimation of the model states and the grid points, which are treated as extended model states, within a proper observer or filter scheme, e.g., the \textit{Extended Kalman Filter} scheme.
Although this is a very elegant approach, it is rather inappropriate for the application considered here, due to the following reasons: 
i) As a consequence of the real-gas model integration, every execution of the overall multi-stage compressor model is relative costly compared to more simple models; and ii) much more grid points are necessary to shape the multidimensional performance maps properly compared to the application in \cite{hoeckerdal2011ekf}.
Considering that every grid point, treated as extended model state within the joint estimation scheme, increases the number of model executions in every estimation step, an unacceptable increase of the computational load arises for the present application.

Therefore, we will present an alternative approach to map adaption that comprises a filter for state and parameter estimation and a coupled map estimation, which will not trigger additional model executions.
For the first issue, a constrained \textit{Unscented Kalman Filter} (UKF, e.g., see \cite{julier2000new,kolas2009constrained}) is applied  (see Section \ref{sec:CUKF}). % to the issue of estimating unmeasurable process states and two bias terms per compressor stage (see Section \ref{sec:CUKF}).
%A constrained \textit{Unscented Kalman Filter}  is applied to the issue of tracking unmeasured or unmeasurable process states quickly .
The UKF is a real-time state estimator for nonlinear systems, applicable even if no Jacobian matrix could be achieved.
In this paper, the term ``real-time'' refers to a situation where the measurement sampling rate is assumed to be low enough to complete all calculation steps between two measurement samples in real time.
For this, all presented algorithms are formulated in an effective, recursive manner.
% thus, they are capable of incorporating measurements as they arise, i.e., in real-time.
%In order to assure compliance with physical constraints, a UKF scheme investigated in \cite{kolas2009constrained} is applied, see Section \ref{sec:CUKF}.
For real-time performance map estimation, an algorithm referred to as \textit{Recursive Map Estimation} (RME) is presented in Section \ref{sec:map_estimation}, and combined with the UKF in Section \ref{sec:CSME}.
The overall monitoring performance is assessed in Section \ref{sec:results} for a simulative experiment with known state trajectories, performance maps, and alterations.
A two-stage supercritical \COO\ compressor acts as the reference process.

% desire for dynamic models and dynamic supervision (transients)
% first-principle approach
% Centrifugal compression for real-gas applications

% Multi-stage --> two-stage here

% Aufbau:
%The methodological part of the document is divided into two parts.

%%%%%%%%%%%%%%%%%%%%%%%%%%%%%%%%%%%%%%%%%%%%%%%%%%%%%%%%%%%%%%%%%%%%%%
%\section{The Scope of the Application}
%\label{sec:scope}

% real-gas
% flexible
% CORA

%$$ T_\mathsf{1}^\mathsf{0} \ T_\mathtt{1}^\mathtt{0} \ T_\mathrm{1}^\mathrm{0} \ T_\mathsf{\mathbf{1}}^\mathsf{\mathbf{0}} \ T_\mathbf{\mathbf{1}}^\mathbf{\mathbf{0}} \ T_1^0 $$

%%%%%%%%%%%%%%%%%%%%%%%%%%%%%%%%%%%%%%%%%%%%%%%%%%%%%%%%%%%%%%%%%%%%%%
\section{Model Building}
\label{sec:model}
% pressure computation (p1,v1,h1,a1,cv1,dp_dT_v1)
% stage 1 -> (rho1,h0,a0)
% (v1,h1,a1) -> stage 2

\subsection{Model Structure and Nomenclature}
\label{sec:structure}

The starting point for the model-based monitoring approach is a nonlinear, dynamic system description
\begin{numcases}{}
\difffrac{\smallfcn{\vec{x}}{t}}{t} = \fcn{\vec{f}}{\vec{x},\,\vec{u},\,\vec{\theta},\, t}\ , 	 & $\smallfcn{\vec{x}}{t_0}=\vec{x}_0$ , \label{eq:ZDGL} \\[.5em]
\hphantom{\difffrac{ \smallfcn{\vec{x}}{t}}{t}} \negphantom{\smallfcn{\vec{y}}{t}}
\smallfcn{\vec{y}}{t} = \fcn{\vec{g}}{\vec{x},\,\vec{u},\,\vec{\theta},\, t}\, ,  & \label{eq:Ausgangsgleichung}
\end{numcases}
where $\vec{y} \in \mathbb{R}^{n_y}$, $\vec{x} \in \mathbb{R}^{n_x}$, $\vec{u} \in \mathbb{R}^{n_u}$, and $\vec{\theta} \in \mathbb{R}^{n_\theta}$ are the measurable outputs, the states, the inputs, and the parameters of the model, respectively.
In general, all of these values are time-variant, but the model parameters are assumed to vary much slower than the other variables.
To increase readability of the equations, the time argument $t$ is suppressed in what follows.

Consider the $N$-stage compression unit depicted in Fig.\ \ref{fig:unit}, with a suction pipe P0 and a discharge pipe PN.
The model states and outputs might be structured as follows:
\begin{align}
	\renewcommand{\arraystretch}{1.25}
	\vec{x} = 		\begin{bmatrix}	 \Pa{\vec{x}} \\	\Sa{\vec{x}} \\ \Pb{\vec{x}} \\ \Sb{\vec{x}} \\ \Pc{\vec{x}} \\ \vdots \\ \SN{\vec{x}} \end{bmatrix} \, , \quad
%	\vec{y} &= 		\begin{bmatrix}	 (\Pa{\vec{y}})^T &	(\Sa{\vec{y}})^T & (\Pb{\vec{y}})^T & (\Sb{\vec{y}})^T & (\Pc{\vec{y}})^T & \dotsm & \SN{T_s}	\end{bmatrix}^T \, ,
	\vec{y} = 		\begin{bmatrix}	 \Pa{\dot{m}} \\ \Pa{\vec{y}} \\	\Sa{T_s} \\ \Pb{\vec{y}} \\ \Sb{T_s} \\ \Pc{\vec{y}} \\ \vdots \\ \SN{T_s}	\end{bmatrix} \, , \label{eq:states}
	\renewcommand{\arraystretch}{1}
\end{align}
where the individual state vectors of a single component
\begin{align}
	\renewcommand{\arraystretch}{1.25}
	\Pj{\vec{x}} = 	\begin{bmatrix}		\Pj{\Tr_f} \\ \Pj{\Tr_s} \\ \Pj{\vr} \end{bmatrix} \, , \quad \Pj{\vec{y}} = 	\begin{bmatrix}		\Pj{T_s} \\ \Pj{p} \end{bmatrix} \, ,  \quad \Si{\vec{x}} = \begin{bmatrix}	\Si{\Tr_f} \\ \Si{\Tr_s} \\ \Si{\Dmu} \\ \Si{\Dphi} 	\end{bmatrix} \, \label{eq:xP0}
	\renewcommand{\arraystretch}{1}	
\end{align}
are detailed below. %, and $\left[~\right]^T$ denotes the transpose of a vector.
The known or measured model inputs are the speeds of the compressor shafts and the discharge pressure:
\begin{align}
	\vec{u} = \begin{bmatrix}	 \Sa{n} & \Sb{n} & \dotsm & \SN{n} & \PN{p}	\end{bmatrix}^T\, . \label{eq:u}
\end{align}
Here, $\left[~\right]^T$ denotes the transpose of a vector.
Model parameters $\vec{\theta}$ result from first-principle modeling, and are specified below when they appear.
The superscripts Pj and Si denote whether the respective physical quantity belongs to the suction pipe (P0), the $i$th compressor stage, the $j$th intermediate pipe, or the discharge pipe (PN).
The physical quantities are the temperatures $T$, pressures $p$, and specific volumes $v$, or their dimensionless counterparts:
\begin{align}
	\Tr &= \frac{T}{T_c}\, , \qquad \pr = \frac{p}{p_c}\, , \qquad \vr = v\,\frac{p_c}{R\, T_c}\, ,
\end{align}
respectively.
For scaling, $R$ is the specific gas constant, and $T_c$ and $p_c$ are the critical temperature and pressure of the process fluid, respectively.
Thus, for $(\Tr>1,\ \pr>1)$, the process fluid is at a supercritical state.
The following sections contain further dimensionless thermodynamic quantities:%, which are also shown in Fraktur typeface:
\begin{align}
	\hr = \frac{h}{R\, T_c}\, , \qquad \ur = \frac{u}{R\, T_c}\, , \qquad \cvr = \frac{c_v}{R}\, .
\end{align}
$h$ is the specific enthalpy, $u$ is the specific internal energy, and $c_v$ is the specific isochoric heat capacity.

Additional variables are mass flows $\md$ and deviations ($\Delta$) from nominal compressor specific quantities $\mu$ and $\varphi$, which are introduced in Section \ref{sec:stagemap}.
For the temperature values modeled, a distinction is drawn between fluid temperatures and temperatures at the sensor location (for the same pipe cross-section), denoted by subscripts $f$ and $s$, respectively.
This distinction is necessary, because temperature sensors are quite often shielded by thick-walled casings, especially for high-pressure applications, leading to considerably delayed measurements.
Temperature values assigned to a compressor stage are discharge temperatures always; for example, $\Sa{T_f}$ and $\Sa{T_s}$ denote the discharge temperatures (the fluid and sensor position) of the first stage.
Elsewhere, if a local assignment in the context of a single plant component (compressor stage or pipe) is needed, entry values are indicated with a subscript $1$, and exit values with a subscript $2$.

\subsection{Calculation of Real-Gas Values}
\label{sec:realgas}

For the sake of adaptability of the approach to fluids other than that considered in the example below, a generic real-gas state equation with a low parametrization effort is chosen.
For a multitude of relevant process fluids (air, hydrocarbons, carbon dioxid, hydrogen, and ammonia), the LKP state equation shows good agreement with published gas property tables \cite{ludtke2004process}.
The equation is based on the \textit{three-parameter corresponding states} principle \cite{lee1975LKP} featuring reduced temperature $\Tr$, reduced pressure $\pr$, and acentric factor $\omega$.
For a given real gas (mixture), its (pseudo-)critical temperature $T_c$ and pressure $p_c$, as well as its acentric factor $\omega$ and its specific gas constant $R$, determine all thermal relations for the LKP model. 
The thermal relations are formulated as follows:
\begin{align}
	&\pr = \fcn{\pr_S}{\Tr,\vr_S} \, , \quad \pr = \fcn{\pr_R}{\Tr,\vr_R}\, ,  \label{eq:p-T-v} \\
	&\text{and} \quad \vr = \fcn{\delta}{\vr_S, \vr_R, \omomR} \, , \label{eq:v_vS_vR} 
\end{align}
where $\pr_S$ and $\pr_R$ are separated \textsc{Benedict-Webb-Rubin-Starling} (BWRS)-type equations for a \textit{simple} fluid and a \textit{reference} fluid, respectively, $\omega_R$ is the acentric factor of the \textit{reference} fluid, and $\vr_S$ and $\vr_R$ are the reduced specific volumes of these fluids, which are used to interpolate the reduced real gas (mixture) specific volume $\vr$ according to
\begin{align}
	\fcn{\delta}{\vr_S,\, \vr_R,\, \omomR} := \vr_S + \omomR \cdot \left[ \vr_R - \vr_S \right]\ . \label{eq:fi}
\end{align}
The common, originally stated way to resolve the thermal relations (\ref{eq:p-T-v})--(\ref{eq:v_vS_vR}) is as follows (see \cite{lee1975LKP}):
Given a thermodynamic state $(\Tr,\pr)$, a typically multiple-step numerical procedure is applied to determine the pair $(\vr_S,\vr_R)$ that fulfills the equality constraint (\ref{eq:p-T-v}).
We refer to this as the \textit{pressure explicit approach}.
At this point, we recommend the method described in \cite{mills1980BWRS}, which is excellent in terms of numerical convergence and reliability in the whole range of valid thermodynamic states\footnote{The LKP state equation is valid for $(\Tr, \pr) \in \left[0.3;\ 4\right] \times \left]0;\ 10\right]$. Due to its continuous pressure explicit formulation, it is improper to describe the multiphase region correctly. The recommended method guarantees an (always existing) solution outside this region.}.
Once $(\vr_S,\vr_R)$ is determined, all thermodynamic properties (e.g., reduced specific enthalpy $\hr$) and their derivatives (e.g., isochoric pressure variation $(\partial \pr / \partial \Tr)_\vr$) can be calculated directly.
Calculations in this \textit{pressure explicit approach} are abbreviated in a respective manner, e.g., $\hr(\Tr,\pr)$ or $(\partial \pr / \partial \Tr)_\vr(\Tr,\pr)$.  %$\pd{\pr}{\Tr}{\vr}\!\!\!(\Tr,\pr)$.
For the sake of completeness, a thermal state equation, such as the LKP state equation, is able to determine the deviation of the ideal-gas caloric properties only.
Thus, caloric ideal-gas data, e.g., the thermal dependence of the isochoric heat capacity $c_v^{id}=c_v^{id}(T)$, is necessary to calculate absolute caloric values.
%All relevant expressions are summarized in the appendices.

%However, for the proposed real-time application such a (multiple-step) procedure comes with an unacceptable numerical burden.
%Therefore, if the range of possible thermodynamic states is restricted to the gaseous and supercritical region, we found an alternative approach, which is much less computationally intensive and sufficiently accurate.
%Finally, it yields to the (one-step solvable) solution:
If the range of possible thermodynamic states is restricted to the gaseous and supercritical regions, and if the thermodynamic state can be determined by $(\Tr,\vr)$, we found an alternative approach, which is less computationally intensive and sufficiently accurate.
Within this approach, given $(\Tr,\vr)$, but without knowing $\pr$ \textit{a priori}, it is clear from (\ref{eq:p-T-v}) that \mbox{$\pr=\pr_S(\Tr,\vr_S)=\pr_R(\Tr,\vr_R)$}, or in an alternative mathematical description, \mbox{$\epsilon(\vr_S,\vr_R)=\pr_S(\Tr,\vr_S)-\pr_R(\Tr,\vr_R)=0$} has to be fulfilled.
Treating formula (\ref{eq:v_vS_vR}) as equality constraint, the problem boils down to a scalar root determination of $\epsilon$ with merely one independent variable, e.g., $\vr_S$.
Applying a second-order root determination scheme, we found that a single step (iteration) results in a sufficiently small $\left|\epsilon\right|$ if the fluid is in a gaseous or supercritical phase and if the starting point of the root determination algorithm is set to $\vr_{S,0}=\vr$.
%Contrary to the \textit{pressure explicit approach}, wherein two independent roots of the residuum functions \mbox{$\epsilon_S(\vr_S)=\pr-\pr_S(\Tr,\vr_S)$} and \mbox{$\epsilon_R(\vr_R)=\pr-\pr_R(\Tr,\vr_R)$} have to be determined with $(\Tr, \pr)$ given, one could search for the root of 
Finally, this approach yields the (one-step solvable) solution:
\begin{align}
	\left(\vr_S,\ \vr_R\right) = \left( \vr_{S,1},\ \delta\left(\vr_{S,1},\vr,\frac{\omega_R}{\omega}\right) \right)\, ,
\end{align}
where
\begin{align}
\vr_{S,1} &= \vr - \frac{ \epsilon' + \sqrt{ \fcn{}{\epsilon'}^2 - 2\, \epsilon\, \epsilon'' } }{ \epsilon'' }\, , \label{eq:vrS1}\\[1em]
\epsilon &=  \fcn{\pr_S}{\Tr,\vr} -  \fcn{\pr_R}{\Tr,\vr}\, , \\
\epsilon' &= \frac{\partial\, \pr_S (\Tr,\vr_S)}{\partial\, \vr_S}\Big|_{\vr_S=\vr} - \left[1-\frac{\omega_R}{\omega}\right]\frac{\partial\, \pr_R (\Tr,\vr_R)}{\partial\, \vr_R}\Big|_{\vr_R=\vr}\, , \\
\epsilon'' &= \frac{\partial^2\, \pr_S (\Tr,\vr_S)}{\partial\, \vr_S^2 }\Big|_{\vr_S=\vr} - \left[1-\frac{\omega_R}{\omega}\right]^2\frac{\partial^2\, \pr_R (\Tr,\vr_R)}{\partial\, \vr_R^2}\Big|_{\vr_R=\vr}\, .
\end{align}
%For a detailed description of the included terms and functions, the reader is referred to the appendices.
Calculations based on the latter approach are abbreviated as $\hr(\Tr,\vr)$ or $(\partial \pr / \partial \Tr)_\vr(\Tr,\vr)$, for example.
We refer to this as the \textit{volume explicit approach}.

%The introduced model below necessitates the computation of further thermodynamic properties, namely, the reduced internal energy $\ur$, the reduced isochoric heat capacity $\cvr$, the speed of sound $a$, and the derivative $\pd{\pr}{\Tr}{\vr}$.
%For reasons that would go beyond the scope of this article, we use 
%\begin{align}
%	\pd{p_r}{T_r}{v_r} &= -\pd{v_r}{T_r}{p_r}\cdot \pd{v_r}{p_r}{T_r}^{-1} \ , \label{eq:dpr_dTr_vr_MG}\\
%	\pd{p_r}{v_r}{T_r} &= \pd{v_r}{p_r}{T_r}^{-1} \ , \label{eq:dpr_dvr_Tr_MG}
%\end{align}
%where
%\begin{align}
%\pd{v_r}{T_r}{p_r}&=  \fcn{f_i}{ \pd{v_r}{T_r}{p_r}^S\Bigg|_{v_r=v_r^S},\, \pd{v_r}{T_r}{p_r}^R\Bigg|_{v_r=v_r^R},\, \omega }\ , \label{eq:dvr_dTr_pr_fi}  \\
%\pd{v_r}{p_r}{T_r}&=  \fcn{f_i}{ \pd{v_r}{p_r}{T_r}^S\Bigg|_{v_r=v_r^S},\, \pd{v_r}{p_r}{T_r}^R\Bigg|_{v_r=v_r^R},\, \omega }\ . \label{eq:dvr_dpr_Tr_fi}
%\end{align}
%instead of $\cvr$,... according to \cite{lee1975LKP}...

\subsection{Compressor Stage Model}
\label{sec:stagemap}

Following the basic design philosophy for (real-gas) centrifugal compressors in \cite{ludtke2004process}, a static dependency between dimensionless characteristic compressor numbers \mbox{$\Psi_p$--$\varphi$--$\mu$--$\Mu$} of a specific compressor stage is postulated.
The dimensionless numbers are
\begin{flalign}
&\text{the polytropic head coefficient} &\Psi_p &= 2\,\frac{y_p}{\tipspeed^2}\,, \label{eq:Psi}\\
&\text{the flow coefficient} & \varphi &= \frac{4}{\pi}\frac{\dot{V}_s}{d_2^2 \tipspeed}\,, \label{eq:phi}\\
&\text{the work input factor} & \mu &= \frac{\Delta h}{\tipspeed^2}\, , &\label{eq:mu}\\
&\text{and the machine Mach number} & \Mu &= \frac{\tipspeed}{a_1}\, . &\label{eq:Mu2}
\end{flalign}
The variables used are listed in Table \ref{tab:characteristics}.
\begin{table}
	\centering
	\renewcommand{\arraystretch}{1.4}
	\begin{tabular}{lcl}
		\hline 
		Polytropic work	& $y_p$ & \SI{}{\J\per\kg} \\ 
	%	\hline 
		Actual compression work & $\Delta h$ & \SI{}{\J\per\kg} \\ 
	%	\hline 
		Impeller diameter & $d_2$ & m \\ 
	%	\hline 
		Blade speed at impeller exit & $\tipspeed$ & \SI{}{\meter\per\second} \\ 
	%	\hline 
		Suction volume flow & $\dot{V}_s$ &  \SI{}{\cubic\meter\per\second} \\ 
	%	\hline 
		Sonic inlet velocity & $a_1$ & \SI{}{\meter\per\second} \\ 
		\hline 
	\end{tabular} 
	\caption{Compressor characteristics and their physical units}
	\label{tab:characteristics}
	\renewcommand{\arraystretch}{1}
\end{table}
With these terms, the polytropic efficiency can be introduced that serves as a meaningful assessment measure, concerning not only the efficiency of the current operating point but also the health status of the respective compressor stage:
\begin{align}
	\eta_p = \frac{y_p}{\Delta h} = \frac{\Psi_p}{2\, \mu}\, . \label{eq:etap}
\end{align}

For real compressors, a generically structured mapping function, e.g., a fixed-order polynomial, is improper for describing the multitude of possible shapes of specific compressor performance maps \mbox{$\Psi_p$--$\varphi$--$\mu$--$\Mu$} accurately.
That is why we present an interpolation-based technique in Section \ref{sec:map_estimation} to approximate arbitrary shapes.
In this section, the actual shape is irrelevant, and we focus on the solution strategy to determine the current operating point of a compressor stage within its performance map.
To this end, one has to ask, ``Which of the properties of $\Psi_p$, $\varphi$, $\mu$, and $\Mu$ are \textit{a priori} accessible and determine the remaining properties uniquely?''
Given the predefined model structure (see Section \ref{sec:structure}), the dynamic simulation will supply the suction and discharge states, as well as the rotational speeds for every time instant. Thus, $\Mu$ and $\Psi_p$ can be calculated readily; see below. As a result, we propose an interpolation-based mapping %\footnote
%{
%The essential attribute `bijective' holds for $\pd{\Psi_p}{\varphi}{\Mu}<0$; in other words, the region of compressor surge is excluded. 
%}
\mbox{$\mathfrak{M}: (\Mu, \Psi_p) \rightarrow (\varphi, \mu)$}, to uniquely determine $\varphi$ and $\mu$.
With this information, the compressor map can be displayed in the conventional way as $\Psi_{p}$ as a function of $\varphi$; see Section \ref{sec:mapMonitor} as well.
The shape of $\mathfrak{M}$ is addressed in Section \ref{sec:map_estimation}.

\begin{figure*}
	\centering
	\includegraphics[width=\textwidth]{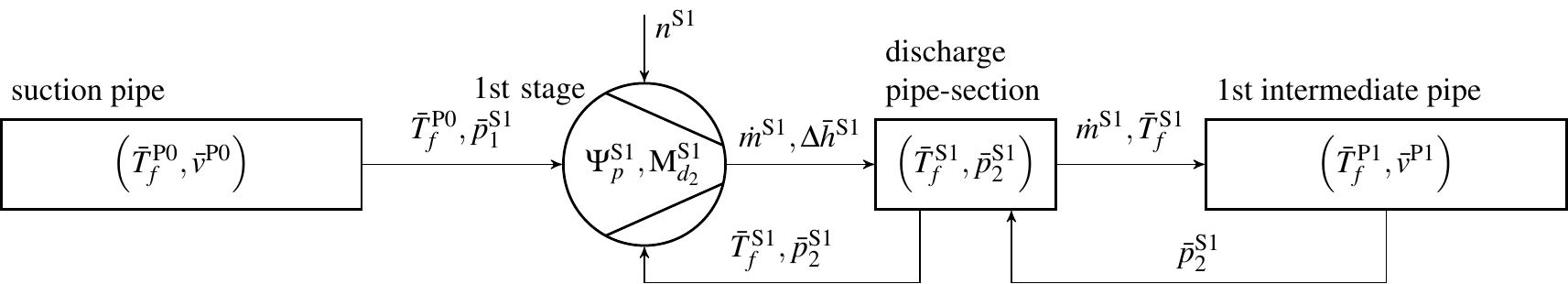}
	\caption[The dynamic compressor stage model concept]{The dynamic compressor stage model concept exemplified for the first stage;\\ in-block values denote current states, values displayed on the paths represent drivers for changing linked in-block states}
	\label{fig:stage model}
\end{figure*}

To determine an operating point within such a \textit{polytropic} performance map, an iterative multiple-step procedure is unavoidable, because it is infeasible to determine the discharge temperature without prior knowledge of the operating point, and vice versa.
Within the framework of the entire monitoring scheme, it turned out to be advantageous to resolve this dependence with the formulation of a dynamic problem.
%A linear black-box model could serve this purpose; however, we found it reasonable to choose a physically motivated approach by formulating 
To this end, an energy balance for an appropriately defined hypothetical volume element featuring the artificial dynamic state variables $\Sj{\Tr_f}$ is formulated as a physically motivated approach; see Eq.\ (\ref{eq:xP0}) as well.
The model should ensure that such temperatures tend rapidly toward values that are consistent with the converged operating point.
As an example for the $j$th stage, the approach results in
\begin{align}
	\difffrac{ \Sj{\Tr_f}	}{ t} = \frac{ \left|\Sj{\dot{m}}\right| \left[\Sj{\hr_1}+\Sj{\Delta\hr} - \Sj{\hr_2}\right] \Sj{\vr_2} }{ \Sj{V_2} \Sj{\cvr_2} }\ \cdot \frac{R\, T_c}{p_c}\ . \label{eq:dTfS_dt}
\end{align}
A steady discharge temperature $\Sj{\Tr_f}$ is obtained if and only if the actual reduced compression work \mbox{$\Sj{\Delta\hr}$}, computed from the performance map, equals the direct reduced enthalpy increase over the compressor stage \mbox{$\Sj{\hr_2} - \Sj{\hr_1}$}, i.e., the calculation is converged.
Within this concept, shown in Fig.\ \ref{fig:stage model} for the first stage, $\Sj{\Tr_f}$ is the temperature, corresponding to the reduced enthalpy $\Sj{\hr_2}$, of the small, fixed-size fluid volume $\Sj{V_2}$, which can be interpreted as a short pipe section, connected to the discharge side of the compressor stage.
For the sake of consistency, this artificial fluid volume is under the same pressure as the actual volume within the downstream pipe.
$\Sj{V_2}/\Pj{V}$ must be small enough to guarantee the desired ``rapid'' solution, but large enough to avoid dynamic stiffness.
This trade-off is addressed within Section \ref{sec:CSME}, where we present an approach, which is proven to handle $\Sj{V_2}/\Pj{V}\approx 0.01$.

Here, for the $j$th compressor stage, the concrete, proposed solution strategy to determine an operating point within the performance map and the right side of Eq.\ (\ref{eq:dTfS_dt}), given the thermodynamic states $(\Pjm{\Tr_f},\Pjm{\vr})$ at the suction side and $(\Pj{\Tr_f},\Pj{\vr})$ at the discharge side, which is an intermediate pipe in the global scheme for $j\ne N$ (cf.\ Fig.\ \ref{fig:unit}), as well as the current discharge temperature $\Sj{\Tr_f}$, is as follows:
\begin{enumerate}[leftmargin=*]
	\item Determine the reduced specific enthalpy, pressure, and speed of sound at the inlet, utilizing the \textit{volume explicit approach} (see Section \ref{sec:realgas}) \label{item:calc_h1}
	\begin{flalign*}
		& \Sj{z_1}=\fcn{z}{\Pjm{\Tr_f},\Pjm{\vr}}\,, \ \text{for}\  z=\{ \hr,\, \pr,\, a \}\, ,&
	\end{flalign*}
	which, inter alia, leads to the machine Mach number $\Sj{\Mu}$ if the given compressor shaft speed $\Sj{n}$ and impeller diameter $\Sj{d_2}$ are taken into account.
	\item Determine the reduced discharge pressure, utilizing the \textit{volume explicit approach} \label{item:calc_p2}
	\begin{flalign*}
		& \Sj{\pr_2}=\fcn{\pr}{\Pj{\Tr_f},\Pj{\vr}}\, .&
	\end{flalign*}		
	\item Determine the reduced specific enthalpy, volume, and isochoric heat capacity at the discharge pipe-section, utilizing the \textit{pressure explicit approach} (see Section \ref{sec:realgas}) \label{item:calc_h2} as $\Sj{\pr_2}$ is given from the last step
%	$$\Sa{\hr_2}=\hr(\Sa{\Tr_f},\Sa{\pr_2})\,, \ \Sa{\vr_2}=\vr(\Sa{\Tr_f},\Sa{\pr_2})\, ,\ \Sa{\cvr_2}=\cvr(\Sa{\Tr_f},\Sa{\pr_2})\, ;$$
	\begin{flalign*}
		& \Sj{z_2}=\fcn{z}{\Sj{\Tr_f},\Sj{\pr_2}}\, , \ \text{for}\ z=\{ \hr,\, \vr,\, \cvr \}\, .&
	\end{flalign*}
	\item Calculate the polytropic volume exponent	\label{item:calc_nv}
	\begin{flalign*}
		& \Sj{n_v} = - \frac{ \ln(\Sj{\pr_2}/\Sj{\pr_1}) }{ \ln(\Sj{\vr_2}/\Pjm{\vr}) }\, .&
	\end{flalign*}
	\item Calculate the polytropic work \label{item:calc_yp}
	\begin{flalign*}
		& \Sj{y_p} = \Sj{\pr_1} \Pjm{\vr} \frac{\Sj{n_v}}{\Sj{n_v}-1} \left[ \left(\frac{\Sj{\pr_2}}{\Sj{\pr_1}}\right)^{\frac{\Sj{n_v}-1}{\Sj{n_v}}} - 1 \right]  R\, T_c \, ,&
	\end{flalign*}
	which leads to the polytropic head coefficient $\Sj{\Psi_p}$.
	\item 	Apply the mapping $\Sj{\mathfrak{M}}: (\Sj{\Mu}, \Sj{\Psi_p}) \rightarrow (\Sj{\varphi}, \Sj{\mu})$ to determine the remaining properties $\Sj{\varphi}$ and $\Sj{\mu}$. \label{item:calc_phi}
\end{enumerate}
Note that steps \ref{item:calc_nv} and \ref{item:calc_yp} are consistent with the definition of a polytropic change $\{p v^{n_v}=\text{const.}\,,\ n_v=\text{const.}\}$ according to \textsc{Zeuner} (cf.\ \cite{zeuner1866grundzuge}), which is occasionally considered an approximation for real gases only, although an approximation of a polytropic change $\{\eta_p=v \, dp/dh\,,\ \eta_p=\text{const.}\}$ according to \textsc{Stodola} (cf.\ \cite{stodola2013dampf}) is actually meant.
However, according to the authors, both definitions are approximations of the real change-of-state path, and choosing \textsc{Zeuner}'s approach is inevitable in the given context, to enable real-time capability.
References \cite[pp.\ 385--386]{baehr1996warme} and \cite{wettstein2014polytropic} are given as indications for further discussion on this topic.

To determine the right side of Eq.\ (\ref{eq:dTfS_dt}), the remaining terms, namely, the mass flow and the reduced compression work, follow from:
\begin{align}
%	\Sj{\md} &= \frac{\Sj{\dot{V}_s}}{\Pjm{v}} =  \frac{\pi^2}{4} \frac{\Sj{\varphi} + \Sj{\Delta \varphi}}{\Pjm{\vr}} \, \Sj{n} \, {\Sj{d_2}}^3 \cdot  \frac{p_c}{R\, T_c}\ , \label{eq:mS1}\\
%	\Sj{\Delta\hr} &= \left(\Sj{\mu} + \Sj{\Delta \mu}\right) \pi^2 \, {\Sj{n}}^2 \, {\Sj{d_2}}^2 \cdot \frac{1}{R\, T_c}\ .
	\Sj{\md} &= \frac{\Sj{\dot{V}_s}}{\Pjm{v}} =  \frac{\Sj{\varphi} + \Sj{\Delta \varphi}}{\Pjm{\vr}} \, \frac{\pi}{4}\, {\Sj{d_2}}^2 \left[\pi\, \Sj{n}\, \Sj{d_2} \right] \cdot  \frac{p_c}{R\, T_c}\ , \label{eq:mS1}\\
	\Sj{\Delta\hr} &= \left[\Sj{\mu} + \Sj{\Delta \mu}\right] \left[\pi \, \Sj{n}\, \Sj{d_2}\right]^2 \cdot \frac{1}{R\, T_c}\ .
\end{align}
Note the incorporation of the model states $\Sj{\Delta \varphi}$ and $\Sj{\Delta \mu}$, which represent deviations from the (nominal) performance map, and which will be estimated below.

For the $N$th compressor stage, the procedure is analogous, with the exception of step \ref{item:calc_p2}, which is obsolete due to the given model input $\PN{p}$ (cf.\ Eq.\ (\ref{eq:u})).
Note that the multi-stage model can be programmed efficiently, because many interim results, e.g., the pressure within the intermediate pipe $\Sjm{\pr_2}=\Sj{\pr_1}$, and terms from the thermal relations that are not shown here appear multiple times.

\subsection{Dynamic State Equations}
\label{sec:state_eq}

In addition to the artificial state equations $\diff{\!\Sj{\Tr_f}}/\diff{\!t}$ (see Eq.\ (\ref{eq:dTfS_dt})), dynamic equations for the remaining states have to be derived.
For quantities, representing the thermodynamic state within intermediate pipes, $\Pj{\Tr}$ and $\Pj{\vr}$ for $j=\{1,\dots,N-1\}$, appropriate balance equations are utilized.
To ensure real-time capability, each fluid volume within such pipes is modeled as an open, well-mixed reservoir.
The change of state for such a reservoir, considering real-gas behavior, and expressed in a differential manner, results in
\begin{align}
	c_v\,  \diff{T} &= \frac{v}{V}\left[ \diff{U} - u\cdot \diff{m} + \pd{u}{v}{T}\cdot v\cdot \diff{m} \right]\, , \label{eq:dT}\\
	\diff{ v} &= -\frac{v^2}{V}\cdot \diff{m}\, . \label{eq:dv}
\end{align}
$V$ is the fixed-size reservoir volume, $U$ is the extensive internal energy, $m$ is the volume's total mass, and further,
\begin{equation}
	\pd{u}{v}{T} = T \pd{p}{T}{v} - p
\end{equation}
is a universal caloric relation (cf.\ \cite[p.\ 140]{baehr1996warme}).
The energy and mass balance, given a single upstream input cross-section (subscript $1$) and a single downstream outlet cross-section (subscript $2$), yield
\begin{align}
	\diff{ U} &= h_1\cdot \diff{m_1} - h\cdot  \diff{m_2} + \diff{Q} \, , \label{eq:dU}\\
	\diff{ m} &= \diff{m_1} - \diff{m_2}\, , \label{eq:dm}
\end{align}
where $dQ$ might be used to model additional energy transfer, e.g., heat transfer in the case of a diabatic pipe casing.
The combination of Eqs.\ (\ref{eq:dT})--(\ref{eq:dm}), expressed with dimensionless quantities, leads to the dynamic evolution equations of the thermodynamic state of an intermediate pipe volume:
%\begin{align}
%	\begin{split}
%		\frac{d \Tr_f}{d t} &= \frac{\vr}{V\, \cvr} \Bigg[ \frac{dQ}{dt}\cdot\frac{1}{R\, T_c} + \md_1\left[\hr_1 - \ur \right] - \md_2 \pr \vr\\
%		&\hspace*{2cm}+ \pd{\ur}{\vr}{\Tr}\left[\md_1-\md_2\right]\vr \Bigg]\cdot \frac{R\, T_c}{p_c}\, ,
%	\end{split}\\
%	\frac{d \vr}{d t} &= -\frac{\vr^2}{V} \left[\md_1 - \md_2\right] \cdot \frac{R\, T_c}{p_c}\, .
%\end{align}
%
%Concretely, considering the aforementioned nomenclature within the global model context, we got:
\begin{align}
	\begin{split}
		\difffrac{ \Pj{\Tr_f}}{ t} = &\frac{\Pj{\vr}}{\Pj{V}\, \cvr\PjTv{f}{} } \Bigg[ \difffrac{\Pj{Q}}{t}\cdot\frac{1}{R\, T_c} + \Sj{\md}\\
		&\hspace*{1.0cm} \cdot \left[\Sj{\hr_2} - \Sjp{\hr_1} \right] +  \Pj{\Tr_f} \pd{\pr}{\Tr}{\vr}\!\!\!\PjTv{f}{}  \\
		&\hspace*{1.0cm} \cdot \left[\Sj{\md}-\Sjp{\md}\right]\Pj{\vr} \Bigg]\cdot \frac{R\, T_c}{p_c}\, ,
	\end{split} \label{eq:dTfP1_dt}\\[0.5em]
	\difffrac{\Pj{\vr}}{ t} = & -\frac{(\Pj{\vr})^2}{\Pj{V}} \left[\Sj{\md} - \Sjp{\md}\right] \cdot \frac{R\, T_c}{p_c}\, ,\label{eq:dvP1_dt}
\end{align}
%where
%\begin{align}
%	\Pb{\cvr}&=\cvr \PbTv{f}{}\, ,\\
%	\Pb{\ur}&=\ur \PbTv{f}{}\, ,\\[0.5em]
%	\begin{split}
%	\pd{\ur}{\vr}{\Tr}^{\!\!\!\text{P1}}&= \pd{\ur}{\vr}{\Tr}\!\!\!\PbTv{f}{} \\
%	&= \Pb{\Tr_f} \pd{\pr}{\Tr}{\vr}\!\!\!\PbTv{f}{} - \Sa{\pr_2} 
%	\end{split}
%\end{align}
%are determined using the \textit{volume explicit approach} (see Section \ref{sec:realgas}).
In Eqs.\ (\ref{eq:dTfP1_dt}) and (\ref{eq:dvP1_dt}), leakage mass flows might be considered, which is not done here.
Note that many terms, including hidden ones such as $\Pj{(\vr_S,\vr_R)}$ (cf.\ Section \ref{sec:realgas}), have already been calculated within the compressor stage model (see Section \ref{sec:stagemap}).

For considering delayed temperature measurements, arbitrary sensor models can be applied.
However, as sensor modeling is not the focus of this article, we choose a simple linear, first-order approach:
\begin{align}
	\difffrac{\Pj{\Tr_s} }{t} = \frac{1}{\Pj{\tau} } \left[ \Pj{\Tr_f} - \Pj{\Tr_s}  \right]\, , \qquad	\difffrac{\Sj{\Tr_s} }{t} = \frac{1}{\Sj{\tau} } \left[ \Sj{\Tr_f} - \Sj{\Tr_s}  \right]\, .\label{eq:dTs_dt} %\quad \text{for}\ Z=\{\text{P0},\,\text{S1},\,\text{P1},\,\text{S2}\} \label{eq:dTs_dt}
\end{align}
$\tau$ denotes time constants, which are the parameters to be adjusted later.

To fully describe the system (\ref{eq:ZDGL})--(\ref{eq:Ausgangsgleichung}), more variables have to be known for which balance equations cannot be formulated.
Most prominently, this applies to the deviation variables $\Delta \Sj{\varphi}$ and $\Delta \Sj{\mu}$ of the performance map.
Within the modeling scheme, they are treated as constants.
The monitoring scheme described in Section \ref{sec:CUKF} will be able to estimate such quantities.
For this purpose, and for applying an \textit{Unscented Kalman Filter}, dynamic equations must be formulated for these ``constants'', namely
\begin{align}
\begin{split}
\difffrac{ \Pa{\Tr_f}}{t} = \difffrac{ \Pa{\vr}}{t} = \difffrac{ \Sj{\Dphi}}{t} = \difffrac{\Sj{\Dmu}}{t} =0\,  .
\end{split} \label{eq:x_const}
\end{align}
The entire dynamic equation set $\vec{f}$ of the multi-stage model (cf.\ Eq.\ (\ref{eq:ZDGL})) consists of Eqs.\ (\ref{eq:dTfS_dt}) and (\ref{eq:dTfP1_dt})--(\ref{eq:x_const}).

\subsection{Output Equations}

Considering the terms already calculated within the compressor stage model (Section \ref{sec:stagemap}), the entire set of output equations $\vec{g}$ of the multi-stage model (cf.\ Eq.\ (\ref{eq:Ausgangsgleichung})) simply comprises:
\begin{align}
\begin{split}
	&\Pa{\md} = \Sa{\md}\, , \quad \Pj{T_s} = \Pj{\Tr_s}\, T_c\, , \quad \Pj{p} = \Sjp{\pr_1}\, p_c\, ,\\
	&\Sj{T_s} = \Sj{\Tr_s}\, T_c\, .
\end{split}
\end{align}
%and $\Sa{\md}$ has already been stated in Eq.\ (\ref{eq:mS1}).

%%%%%%%%%%%%%%%%%%%%%%%%%%%%%%%%%%%%%%%%%%%%%%%%%%%%%%%%%%%%%%%%%%%%%%
\section{Monitoring}
\label{sec:monitoring}

\subsection{Constrained Unscented Kalman Filter}
\label{sec:CUKF}
%Since the aim is to focus on the novelty of our research and to point out the beyond-state-of-the-art modeling and performance map estimation, no detailed introduction to the \textit{Unscented Kalman Filter} (UKF) is given, which is a fundamental part of the overall monitoring scheme.
Because the aim is to focus on the novelty of our contribution, i.e., the modeling approach and the performance map estimation below, the reader is referred to \cite{julier1997new,julier2000new} for an introduction to the well-known UKF.
However, as the UKF is a fundamental part of the overall monitoring scheme, some remarks are in order.
The objective of the UKF approach is to provide an estimate $\vec{\hat{x}}$ of the true, partly unmeasurable states $\vec{x}$ of a nonlinear system that is given by Eqs.\ (\ref{eq:ZDGL}) and (\ref{eq:Ausgangsgleichung}).
%\begin{numcases}{}
%\difffrac{\smallfcn{\vec{x}}{t}}{t} = \fcn{\vec{f}}{\vec{x},\,\vec{u},\,\vec{\theta},\, t}\ , 	 & $\smallfcn{\vec{x}}{t_0}=\vec{x}_0$ , \label{eq:ZDGL_II} \\[.5em]
%\hphantom{\difffrac{ \smallfcn{\vec{x}}{t}}{t}} \negphantom{\smallfcn{\vec{y}}{t}}
%\smallfcn{\vec{y}}{t} = \fcn{\vec{g}}{\vec{x},\,\vec{u},\,\vec{\theta},\, t}\, ,  & \label{eq:Ausgangsgleichung_II}
%\end{numcases}
%where $\vec{y} \in \mathbb{R}^{n_y}$, $\vec{x} \in \mathbb{R}^{n_x}$, $\vec{u} \in \mathbb{R}^{n_u}$, and $\vec{\theta} \in \mathbb{R}^{n_\theta}$ are the measurable outputs, the states, the inputs, and the parameters of the model, respectively.
Readers who are more interested in the basics of implementing this method might prefer the brief presentation in \cite{merwe2001squareroot}.

There are several concepts of the UKF that differ in detail; cf.\ \cite{wu2005unscented,kolas2009constrained,julier2002scaled,sarkka2007unscented}.
For the multi-stage monitoring scheme of a compressor considered here, the \textit{Constrained Unscented Kalman Filter} (CUKF) with the reformulated correction step proposed in \cite{kolas2009constrained} is combined with an additive noise assumption.
%Therefore, the following stochastic, nonlinear, discrete-time system description is derived from Eqs.\ (\ref{eq:ZDGL_II})--(\ref{eq:Ausgangsgleichung_II}):
Therefore, the following stochastic, nonlinear, discrete-time system description is derived from Eqs.\ (\ref{eq:ZDGL}) and (\ref{eq:Ausgangsgleichung}):
\begin{numcases}{\hspace*{-7mm}}
	\xk = \fcn{\vec{F}}{\xkm,\, \vec{u}_{k-1},\, \vec{\theta},\,  k} + \zkm{r}^x\, , \hspace*{-2mm} & $\vec{x}_0$ -- given,\\[.5em]
	\hphantom{\xk} \negphantom{\zk{y}}
	\zk{y} = \fcn{\vec{G}}{\xk,\, \vec{u}_{k-1},\, \vec{\theta},\,  k} + \zk{r}^y\, .  & 
\end{numcases}
A variable with an index $k$ denotes a discrete-time quantity; e.g., $\zk{z}$ would be an abbreviation for a time-sampled value \mbox{$\vec{z}(t=t_k)$} (usually, \mbox{$t_k=k \times \text{``fixed sample time''}$}, \mbox{$k \in \mathbb{N}$}), $\vec{r}^x$ is an additive system noise, and $\vec{r}^y$  represents measurement noise.
$\vec{r}_k^x$ and $\vec{r}_k^y$ are stochastic, zero-mean, uncorrelated, discrete signals with time-variant covariance matrices $\mat{R}_k^x$ and $\mat{R}_k^y$, respectively.
Applying the expectation operator $E\{\}$,
\begin{align}
	\begin{split}
		&\fcnII{E}{\vec{r}_i^x (\vec{r}_j^x)^T} = \mat{R}_k^x\, \delta_{ij} \, , \quad \fcnII{E}{\vec{r}_i^y (\vec{r}_j^y)^T} = \mat{R}_k^y\, \delta_{ij}\, , \\
		&\fcnII{E}{\vec{r}_i^x (\vec{r}_j^y)^T} = \mat{0} \, , \quad \forall\ i,\,j\, ,
	\end{split}
\end{align}
follows, where $\delta_{ii}=1$ and $\delta_{ij}=0$ for $i\ne j$.
%Furthermore, the introduced state update function $\vec{F}$ and the output function $\vec{G}$ are
%\begin{align}
%	\fcn{\vec{F}}{\xkm,\, k} &= \xkm + \int\limits_{t_{k-1}}^{t_k}  \fcn{\vec{f}}{\vec{x},\,\vec{u},\,\vec{\theta},\, t}\, dt \, , \\
%	\fcn{\vec{G}}{\xk,\, k} &= \fcn{\vec{g}}{\vec{x},\,\vec{u},\,\vec{\theta},\, t_k}\, .
%\end{align}

The core of the UKF is the \textit{Unscented Transformation} (UT).
The UT gives an estimate of statistical moments, inter alia, the mean and the covariance, of a density function that is the outcome of a nonlinear transformation (via $\vec{F}$ or $\vec{G}$) of a prior density function.
The estimate is based on specific representatives of the density function, called sigma points.
The sigma points of a distribution of $\vec{x}$ or $\vec{y}$ are typically denoted with $\chi$ or $\gamma$, respectively.

For the multi-stage compressor introduced, physical constraints must be respected.
The valid domains are given in Table \ref{tab:constraints}.
Different types of constraint handling within the \textit{Kalman Filter} approach are discussed in \cite{simon2010constraints}.
In this contribution, a simple approach is chosen:
Check that every sigma point calculated within the UT and the reformulated correction step is in a valid physical domain, and if not, place the entries involved on the nearest element inside the valid domain.
This ad-hoc approach, called \textit{clipping}, is an essential element of the entire monitoring algorithm, because it has a superior stabilizing effect, compared to any internal constraint handling within the general model $\vec{F}$ and $\vec{G}$.
\begin{table}
	\centering
	\renewcommand{\arraystretch}{1.4}
	{\setlength{\tabcolsep}{3pt}
	\begin{tabular}{ll}
		\hline 
		Entry in $\chi$	represents& Valid domain, such that \\
		\hline
		temperature $\Tr$ & $\max (0.3, \Tr_l)  \le \Tr\le \min (4, \Tr_u)$\\ 
		%	\hline 
		specific volume $\vr$ & $\frac{1}{11.7} \le \vr$ \\ 
		%	\hline 
		work input deviation $\Delta \mu$&  $0\le \mu$ and $\eta_p=\frac{\Psi_p}{2\, \mu}\le 1$\\ 
		%	\hline 
		flow deviation $\Delta \varphi$ & $0\le \varphi$ \\
		\hline 
	\end{tabular} 
	\caption[Sigma point constraints]{Sigma point constraints; $\Tr_l$ and $\Tr_u$ represent any known lower and upper temperature bounds for the underlying process}
	\label{tab:constraints}
	\renewcommand{\arraystretch}{1}}
\end{table}

\subsection{Recursive Map Estimation}
\label{sec:map_estimation}

In Section \ref{sec:stagemap}, a generic mapping function \mbox{$\mathfrak{M}: (\Mu, \Psi_p) \rightarrow (\varphi, \mu)$} was introduced to determine an operating point within the dimensionless performance map \mbox{$\Psi_p$--$\varphi$--$\mu$--$\Mu$} of a single compressor stage.
The aim of this work is an estimation of this performance map, even when it changes over time, e.g., due to fouling, or when it is completely unknown from the beginning.
Before going into details, the general idea of the RME is sketched in Fig.\ \ref{fig:RMEsketch}:
\begin{figure*}
	\centering
	\includegraphics[width=\textwidth]{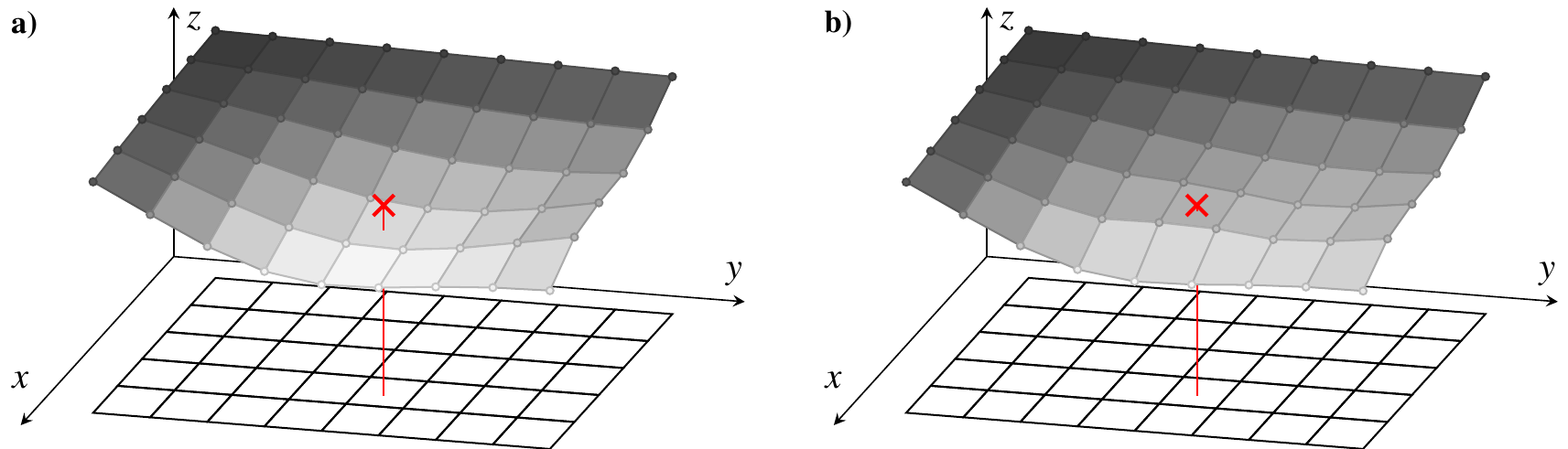}
	\caption[RME sketch]{Scheme of local map adaption;\\ a) the new information \raisebox{0pt}{\tikz{\draw[line width=1.2pt,-,scale=0.6,color=red](0em,0em)--(1em,1em) (1em,0em)--(0em,1em){};}} deviates considerably from the current map; b) the map is adapted by incorporating the new information}
	\label{fig:RMEsketch}
\end{figure*}

Assume that at time $t$ an estimate of the map exists, as is displayed in Fig.\ \ref{fig:RMEsketch}a, for an arbitrary map with two independent variables and one dependent variable.
For $t=0$, the initial guess might be a nominal map or just a horizontal plane $z(x,y)=\text{constant}$.
Data of the actual map is stored for individual pairs of the independent variables $x$ and $y$ on a rectangular grid, as shown as well.
By interpolation, $z$ can be calculated for every pair $(x, y)$.
Now, at time $t$, with the help of the CUKF, an estimate of the process state is obtained that can be exploited to calculate an estimate of the local dependent variable $z(t)$ marked by \raisebox{0pt}{\tikz{\draw[line width=1.2pt,-,scale=0.6,color=red](0em,0em)--(1em,1em) (1em,0em)--(0em,1em){};}} in Fig.\ \ref{fig:RMEsketch}.
This estimate will be used in the RME to adapt the dependent variable $z$ of the map in an optimal manner, in which neighboring $z$-grid values will be more affected than distant ones, and $x$-$y$-grid values will remain in their initial position, as depicted in Fig.\ \ref{fig:RMEsketch}b.
By this, the shape of a time-invariant map can be learned, or a time-variant map can be estimated.

In this sense, for allowing almost arbitrary shapes, we define a performance map via a set of  $N_\mathfrak{M}$ discrete operating points that are initialized for $t=0$, and updated for all future time instants.
The corresponding coordinates $\text{M}_{{d_2},i}^\mathfrak{M}$, $\Psi_{p,i}^\mathfrak{M}$, $\varphi_i^\mathfrak{M}$, and $\mu_i^\mathfrak{M}$ for these operating points $i$ are captured in respective column vectors  $\vec{\text{M}}_{d_2}^{\mathfrak{M}}$, $\vec{\Psi}_{p}^{\mathfrak{M}}$, $\vec{\varphi}^{\mathfrak{M}}$, and \mbox{$\vec{\mu}^{\mathfrak{M}} \in \mathbb{R}^{N_\mathfrak{M}}$}.
Using an interpolation scheme to merge the set of grid points into a coherent map, the mapping function boils down to: %\footnote
%{
%	At this point, further dependencies, e.g., for potential inlet guide vanes, might be formulated (\mbox{$\smallfcn{\vec{m}^T\!\!\!}{ \vec{\text{M}}_{d_2}^{\mathfrak{M}}, \vec{\Psi}_{p}^{\mathfrak{M}}, \vec{z}^\mathfrak{M}, \dots ,\Mu, \Psi_p, z,\dots}$}).
%}
\begin{numcases}{}
	\varphi	=	\fcn{\vec{m}^T\!\!\!}{ \vec{\text{M}}_{d_2}^{\mathfrak{M}}, \vec{\Psi}_{p}^{\mathfrak{M}}, \Mu, \Psi_p}	\cdot	\vec{\varphi}^{\mathfrak{M}}  \, , & \label{eq:map_phi}\\[0.5em]
	\hphantom{\varphi} \negphantom{\mu}
	\mu		=	\fcn{\vec{m}^T\!\!\!}{ \vec{\text{M}}_{d_2}^{\mathfrak{M}}, \vec{\Psi}_{p}^{\mathfrak{M}}, \Mu, \Psi_p}	\cdot	\vec{\mu}^{\mathfrak{M}}  \, , & \label{eq:map_mu}
\end{numcases}
%\begin{align}
%	\begin{bmatrix}
%		\varphi \\ \mu
%	\end{bmatrix}
%	=
%	\begin{bmatrix}
%		\fcn{\mat{M}}{\Mu, \Psi_p} & \fcn{\mat{M}}{\Mu, \Psi_p}
%	\end{bmatrix}
%	\cdot
%	\begin{bmatrix}
%		\vec{\varphi}_{\mathfrak{M}} \\ 		\vec{\mu}_{\mathfrak{M}}
%	\end{bmatrix}
%\end{align}
where %the column vectors $\vec{\text{M}}_{d_2}^{\mathfrak{M}}$, $\vec{\Psi}_{p}^{\mathfrak{M}}$, $\vec{\varphi}^{\mathfrak{M}}$, $\vec{\mu}^{\mathfrak{M}}$ $\in$ $\mathbb{R}^{N_\mathfrak{M}}$ contain the corresponding values $(\Mu)_i^\mathfrak{M}$, $(\Psi_p)_i^\mathfrak{M}$, $\varphi_i^\mathfrak{M}$, $\mu_i^\mathfrak{M}$ of $N_\mathfrak{M}$ grid points of the entire interpolation grid and 
$\vec{m}^T \in \mathbb{R}^{N_\mathfrak{M}}$ is a row vector containing interpolation coefficients that depend on the point to be interpolated $(\Mu,\Psi_p)$ and (usually a subset of) grid points $\vec{\text{M}}_{d_2}^{\mathfrak{M}}$ and $\vec{\Psi}_{p}^{\mathfrak{M}}$. %\footnote
%{
%	As conceptual support, think of a scalar, linear interpolation between three grid points with x-coordinates \mbox{$\vec{x}^\mathfrak{M}=\left[\begin{smallmatrix} x_1^\mathfrak{M} & x_2^\mathfrak{M} & x_3^\mathfrak{M} \end{smallmatrix}\right]^T$} and y-coordinates \mbox{$\vec{y}^\mathfrak{M}=\left[\begin{smallmatrix} y_1^\mathfrak{M} & y_2^\mathfrak{M} & y_3^\mathfrak{M} \end{smallmatrix}\right]^T$} as $$y = \fcn{\vec{m}^T\!\!\!}{ \vec{x}^{\mathfrak{M}}, x}\cdot \vec{y}^{\mathfrak{M}} = \begin{bmatrix} \left[1-\frac{x - x_1^\mathfrak{M}}{x_2^\mathfrak{M} - x_1^\mathfrak{M}}\right] & \frac{x - x_1^\mathfrak{M}}{x_2^\mathfrak{M} - x_1^\mathfrak{M}} & 0 \end{bmatrix} \cdot \begin{bmatrix} y_1^\mathfrak{M} \\ y_2^\mathfrak{M} \\ y_3^\mathfrak{M} \end{bmatrix}\, $$ for $x \in \left[ x_1^\mathfrak{M};\, x_2^\mathfrak{M} \right]$.
%}
Further dependencies, e.g., describing the effect of potential inlet guide vanes, might be included as well. %(\mbox{$\vec{m}^T( \vec{\text{M}}_{d_2}^{\mathfrak{M}}, \vec{\Psi}_{p}^{\mathfrak{M}}, \vec{\alpha}^\mathfrak{M}, \vec{\beta}^\mathfrak{M}, \dots ,\Mu, \Psi_p, \alpha, \beta, \dots)$}).
Note that the structure of $\vec{m}^T$ depends on the selected interpolation method.
The proposed RME is restricted to interpolation methods, where $\vec{m}_k^T$ is not a function of the dependent variables, $\vec{\varphi}^\mathfrak{M}$ or $\vec{\mu}^\mathfrak{M}$.
%We do not aim to derive the applied algorithm that composes $\vec{m}^T$, here.
We utilize a bilinear interpolation method based on a rectangular, normalized interpolation grid to enable efficient programming.

\begin{figure}
	\centering
	\includegraphics{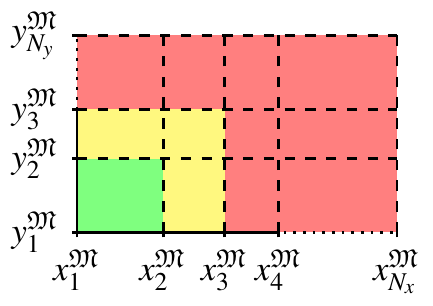}
	\caption[Rectangular grid]{Rectangular grid; the significance of the colors is given in the text}
	\label{fig:grid}
\end{figure}

To exemplify the RME, we return to the $x$-$y$-$z$ notation from the beginning of this section.
Consider a mapping function \mbox{$z	= \smallfcn{ \vec{m}^T }{ \vec{x}^\mathfrak{M}, \vec{y}^\mathfrak{M}, x, y } \cdot \vec{z}^\mathfrak{M}$}, which represents one of the expressions (\ref{eq:map_phi}) or (\ref{eq:map_mu}), with a rectangular grid, as shown in Fig.\ \ref{fig:grid}.
The objective is to estimate the performance map by an optimal adjustment of $\vec{z}^\mathfrak{M}$ considering any (new) information collected. 
The grid vectors are arranged as follows:
\begin{align}
	\begin{split}
		&\kbordermatrix{
		 		& y_1^\mathfrak{M} 		& y_2^\mathfrak{M} 		&		& y_{N_y}^\mathfrak{M} 	\\
		x_1^\mathfrak{M} 	& z_{11} 	& z_{12}	& \dotsm& z_{1N_y} 	\\
		x_2^\mathfrak{M} 	& z_{21} 	& z_{22}	& \dotsm& z_{2N_y} 	\\
		 		& \vdots 	&  			& \ddots&   \vdots	\\
		x_{N_x}^\mathfrak{M} & z_{{N_x}1}& z_{{N_x}2}& \dotsm& z_{{N_x}N_y} \\
		}
		=
		\begin{bmatrix}
			{\vec{z}_1^y}^T \\ {\vec{z}_2^y}^T \\ \vdots \\ {\vec{z}_{N_x}^y}^T
		\end{bmatrix} \\
		&\hspace*{7mm}=
		\begin{bmatrix}
			~{\vec{z}_1^x}~~ & ~{\vec{z}_2^x}~ & \dotsm & ~~{\vec{z}_{N_y}^x}~
		\end{bmatrix} \, .
%		&\vec{z}^\mathfrak{M}
%		=
%		\begin{bmatrix}
%			{\vec{z}_1^x} \\ {\vec{z}_2^x} \\ \vdots \\ {\vec{z}_{N_y}^x}
%		\end{bmatrix}
	\end{split}
\end{align}
With the introduced notation, it is easy to see that
\begin{align}
	\underbrace{
		\begin{bmatrix}
			z_{1j} \\ z_{2j} \\ \vdots \\ z_{{N_x}j}
		\end{bmatrix}
	}_{\displaystyle \vec{z}_j^x} &=
	\underbrace{
		\begin{bmatrix}
			\smallfcn{ \vec{m}^T }{ \vec{x}^\mathfrak{M}, \vec{y}^\mathfrak{M}, x_1^\mathfrak{M}, y_j^\mathfrak{M} } \\
			\smallfcn{ \vec{m}^T }{ \vec{x}^\mathfrak{M}, \vec{y}^\mathfrak{M}, x_2^\mathfrak{M}, y_j^\mathfrak{M} } \\		
			\vdots \\
			\smallfcn{ \vec{m}^T }{ \vec{x}^\mathfrak{M}, \vec{y}^\mathfrak{M}, x_{N_x}^\mathfrak{M}, y_j^\mathfrak{M} }
		\end{bmatrix}
	}_{\displaystyle \mat{M}_{y_j^\mathfrak{M}}}
	\cdot
	\underbrace{
		\begin{bmatrix}
			{\vec{z}_1^x} \\ {\vec{z}_2^x} \\ \vdots \\ {\vec{z}_{N_y}^x}
		\end{bmatrix}
	}_{\displaystyle \vec{z}^\mathfrak{M}} \, ,\\
	\underbrace{
		\begin{bmatrix}
		z_{i1} \\ z_{i2} \\ \vdots \\ z_{i{N_y}}
		\end{bmatrix}
	}_{\displaystyle \vec{z}_i^y} &=
	\underbrace{
		\begin{bmatrix}
		\smallfcn{ \vec{m}^T }{ \vec{x}^\mathfrak{M}, \vec{y}^\mathfrak{M}, x_i^\mathfrak{M}, y_1^\mathfrak{M} } \\
		\smallfcn{ \vec{m}^T }{ \vec{x}^\mathfrak{M}, \vec{y}^\mathfrak{M}, x_i^\mathfrak{M}, y_2^\mathfrak{M} } \\		
		\vdots \\
		\smallfcn{ \vec{m}^T }{ \vec{x}^\mathfrak{M}, \vec{y}^\mathfrak{M}, x_i^\mathfrak{M}, y_{N_y}^\mathfrak{M} }
		\end{bmatrix}
	}_{\displaystyle \mat{M}_{x_i^\mathfrak{M}}}
	\cdot
	\underbrace{
		\begin{bmatrix}
		{\vec{z}_1^x} \\ {\vec{z}_2^x} \\ \vdots \\ {\vec{z}_{N_y}^x}
		\end{bmatrix}
	}_{\displaystyle \vec{z}^\mathfrak{M}} \, \label{eq:My}
\end{align}
is true.
After $k$ time instants, a respective amount of information $\{z_1,z_2,\dots,z_k\}$ (usually, from measurements; below, from CUKF estimates) has been collected, corresponding to $k$ independent operating points $\{(x_1,y_1),(x_2,y_2), \dots,(x_k,y_k)\}$.
The information collected, stored in vector $\vec{z}_k$, will now be used to adapt the $z$-grid values, which are stored in the time-variant vector $\vec{z}_k^\mathfrak{M}$.
It follows:
\begin{align}
	\underbrace{
		\begin{bmatrix}
		z_{1} \\ z_{2} \\ \vdots \\ z_{k}
		\end{bmatrix}
	}_{\displaystyle \vec{z}_k} &=
	\underbrace{
		\begin{bmatrix}
		\smallfcn{ \vec{m}^T }{ \vec{x}^\mathfrak{M}, \vec{y}^\mathfrak{M}, x_1, y_1 } \\
		\smallfcn{ \vec{m}^T }{ \vec{x}^\mathfrak{M}, \vec{y}^\mathfrak{M}, x_2, y_2 } \\		
		\vdots \\
		\smallfcn{ \vec{m}^T }{ \vec{x}^\mathfrak{M}, \vec{y}^\mathfrak{M}, x_k, y_k }
		\end{bmatrix}
	}_{\displaystyle \mat{M}_k}
	\cdot
	\underbrace{
		\begin{bmatrix}
		{\vec{z}_{1,k}^x} \\ {\vec{z}_{2,k}^x} \\ \vdots \\ {\vec{z}_{{N_y},k}^x}
		\end{bmatrix}
	}_{\displaystyle \vec{z}_k^\mathfrak{M}}  
	+ 
	\underbrace{
		\begin{bmatrix}
		e_1^z \\ e_2^z \\ \vdots \\ e_k^z
		\end{bmatrix}
	}_{\displaystyle \vec{e}_k^z}  \, . \label{eq:zk}
\end{align}
%Here, subscript $k$ denote a time stamp for potentially time-variant values.
%Having notified this, the approach pursued can be clarified.
The objective is to minimize the weighted sum of squared errors ${\vec{e}_k^z}^T \mat{W}_{1,k}^z \vec{e}_k^z$, where
\begin{align}
	\vec{e}^z_k = \vec{z}_k - \mat{M}_k \cdot \vec{z}^\mathfrak{M}_k = \vec{z}_k - \mat{M}_k \cdot \left[ \vec{z}_*^\mathfrak{M} + \Delta\vec{z}^\mathfrak{M}_k \right]\, . \label{eq:e_map}
\end{align}
%Remember, $z$ represents $\varphi$ or $\mu$, separately, and, possibly, prior knowledge of a performance map $\vec{z}_*^\mathfrak{M}$ might be incorporated.
If prior knowledge concerning the map is available, this is stored in $\vec{z}_*^\mathfrak{M}$.
Consequently, $\vec{z}_*^\mathfrak{M}$ is fixed, and $\Delta\vec{z}^\mathfrak{M}_k$ is the actual design variable of the optimization problem, the cost function of which is:
\begin{align}
	\begin{split}
		\fcn{J_{1,k}^z}{\Delta\vec{z}_k^\mathfrak{M}} &= {\Delta\vec{z}_k^\mathfrak{M}}^T \mat{M}_k^T \mat{W}_{1,k}^z \mat{M}_k \Delta\vec{z}_k^\mathfrak{M} \\
		&\hspace*{0.3cm}+ 2 \left[ {\vec{z}_*^\mathfrak{M}}^T \mat{M}_k^T - \vec{z}_k^T \right] \mat{W}_{1,k} \mat{M}_k \Delta\vec{z}_k^\mathfrak{M}\, .
	\end{split}	
\end{align}
A unique minimum of $J_{1,k}^z$ exists under very strict conditions only.
%Why is that?
To stress this issue, consider a situation, where all information collected relates to the green area within the grid space, shown in Fig.\ \ref{fig:grid}.
A variation of $z$-values corresponding to grid points within the red area does not affect the interpolated $z$-surface within the green area, and, thus, it has no impact on $\vec{e}^z_k$ and $J_{1,k}^z$, consequently.\footnote
{
	To be precise, the illustration applies only to a C\textsuperscript{0} continuous interpolation method, e.g., a bilinear interpolation, but the issue still exists---to a lesser extent---for C\textsuperscript{n} continuous interpolation functions. 
}
Within the joint estimation scheme proposed by \textsc{H\"ockerdal} et al., see \cite{hoeckerdal2011ekf}, the explained issue transforms into the loss of observability of the respective grid point ``states''.
To overcome this issue, they suggest a specific restriction to the estimated covariance matrix of their approach, with the intention to prevent the divergence of the filter.
The countermeasure here is, to extend the cost function by several regularization terms that add
%\flushleft
%{
\begin{itemize}

	\item[i)] 	the cost of deviations to the \textit{a priori} map ($\vec{z}_*^\mathfrak{M}$)
				\begin{flalign*}
					&	\fcn{J_{2}^z}{\Delta\vec{z}_k^\mathfrak{M}}  = {\Delta\vec{z}_k^\mathfrak{M}}^T \mat{W}_{2}^z \Delta\vec{z}_k^\mathfrak{M} \,  , &
				\end{flalign*}
	\item[ii)] 	the cost of a mean gradient of the entire performance map ($\vec{z}_*^\mathfrak{M} + \Delta\vec{z}_k^\mathfrak{M}$)
				\begin{flalign*}
					&	\fcn{J_{3}^z}{\Delta\vec{z}_k^\mathfrak{M}} = {\Delta\vec{z}_k^\mathfrak{M}}^T  \mat{L}_g^z \Delta\vec{z}_k^\mathfrak{M} + 2\, {\vec{z}_*^\mathfrak{M}}^T \mat{L}_g^z \Delta\vec{z}_k^\mathfrak{M} \, ,&
				\end{flalign*}
	\item[iii)] and the cost of a mean curvature of the entire performance map ($\vec{z}_*^\mathfrak{M} + \Delta\vec{z}_k^\mathfrak{M}$)
				\begin{flalign*}
					&	\fcn{J_{4}^z}{\Delta\vec{z}_k^\mathfrak{M}} = {\Delta\vec{z}_k^\mathfrak{M}}^T  \mat{L}_c^z \Delta\vec{z}_k^\mathfrak{M} + 2\, {\vec{z}_*^\mathfrak{M}}^T \mat{L}_c^z \Delta\vec{z}_k^\mathfrak{M} \, . &
				\end{flalign*}
\end{itemize}
%}
Costs (ii) and (iii) provide the smooth shape of a characteristic performance map.
Matrices $\mat{L}_g^z$ and $\mat{L}_c^z$ are computed as follows:
\begin{align}
	\begin{split}
		\mat{L}_g^z &= \frac{1}{N_y-1} \sum\limits_{j=1}^{N_y-1}  \frac{\Delta \mat{M}_{y_{j}^\mathfrak{M}}^T}{\Delta y_{j}^\mathfrak{M}} \mat{W}_{g_y}^z \frac{\Delta \mat{M}_{y_{j}^\mathfrak{M}}}{\Delta y_{j}^\mathfrak{M}}  \\
		&\hspace*{0.3cm} + \frac{1}{N_x-1} \sum\limits_{i=1}^{N_x-1} \frac{\Delta \mat{M}_{x_{i}^\mathfrak{M}}^T}{\Delta x_{i}^\mathfrak{M}} \mat{W}_{g_x}^z \frac{\Delta \mat{M}_{x_{i}^\mathfrak{M}}}{\Delta x_{i}^\mathfrak{M}} \, ,
	\end{split} 
	\\
	\begin{split}
		\mat{L}_c^z &= \frac{1}{N_y-2} \cdot \sum\limits_{j=1}^{N_y-2}  \frac{\left[ \frac{\Delta \mat{M}_{y_{j+1}^\mathfrak{M}}}{\Delta y_{j+1}^\mathfrak{M}} - \frac{\Delta \mat{M}_{y_{j}^\mathfrak{M}}}{\Delta y_j^\mathfrak{M}} \right]^T}{\frac{y_{j+2}^\mathfrak{M} - y_{j}^\mathfrak{M}}{2}} \mat{W}_{c_y}^z \\
		&\hspace*{0.3cm}\cdot \frac{ \frac{\Delta \mat{M}_{y_{j+1}^\mathfrak{M}}}{\Delta y_{j+1}^\mathfrak{M}} - \frac{\Delta \mat{M}_{y_{j}^\mathfrak{M}}}{\Delta y_j^\mathfrak{M}} }{\frac{y_{j+2}^\mathfrak{M} - y_{j}^\mathfrak{M}}{2}}  + \frac{1}{N_x-2} \\
		&\hspace*{0.3cm} \cdot \sum\limits_{i=1}^{N_x-2}  \frac{\left[ \frac{\Delta \mat{M}_{x_{i+1}^\mathfrak{M}}}{\Delta x_{i+1}^\mathfrak{M}} - \frac{\Delta \mat{M}_{x_{i}^\mathfrak{M}}}{\Delta x_i^\mathfrak{M}} \right]^T}{\frac{x_{i+2}^\mathfrak{M} - x_{i}^\mathfrak{M}}{2}} \mat{W}_{c_x}^z  \frac{ \frac{\Delta \mat{M}_{x_{i+1}^\mathfrak{M}}}{\Delta x_{i+1}^\mathfrak{M}} - \frac{\Delta \mat{M}_{x_{i}^\mathfrak{M}}}{\Delta x_i^\mathfrak{M}} }{\frac{x_{i+2}^\mathfrak{M} - x_{i}^\mathfrak{M}}{2}} \, ,
	\end{split}
\end{align}
where \mbox{$\Delta \mat{M}_{y_j^\mathfrak{M}} = \mat{M}_{y_{j+1}^\mathfrak{M}} - \mat{M}_{y_j^\mathfrak{M}}$} (cf.\ Eq.\ \ref{eq:My}) and \mbox{$\Delta y_j^\mathfrak{M} = y_{j+1}^\mathfrak{M} - y_j^\mathfrak{M}$}, and $\Delta \mat{M}_{x_i^\mathfrak{M}}$ and $\Delta x_i^\mathfrak{M}$ are defined analogously.
All introduced weighting matrices $\mat{W}$ are symmetric and positive definite.
Note that $\mat{L}_g$ and $\mat{L}_c$ are time-invariant, and thus, they can be computed offline.

For the purpose of real-time estimation, a recursive algorithm can be derived to solve the final optimization problem
%\begin{align}
%	\min\limits_{\Delta \vec{z}^\mathfrak{M}_k } \ \smallfcn{J_{1,k}}{\Delta\vec{z}_k^\mathfrak{M}} + \smallfcn{J_{2}}{\Delta\vec{z}_k^\mathfrak{M}} + \smallfcn{J_{3}}{\Delta\vec{z}_k^\mathfrak{M}} + \smallfcn{J_{4}}{\Delta\vec{z}_k^\mathfrak{M}}
%\end{align}
\begin{align}
	\begin{split}
		\min\limits_{ \vec{z}^\mathfrak{M}_k } \ \Big[ &\smallfcn{J_{1,k}^z}{\vec{z}_k^\mathfrak{M}-\vec{z}_*^\mathfrak{M}} + \smallfcn{J_{2}^z}{\vec{z}_k^\mathfrak{M}-\vec{z}_*^\mathfrak{M}} \\
		&+ \smallfcn{J_{3}^z}{\vec{z}_k^\mathfrak{M}-\vec{z}_*^\mathfrak{M}} + \smallfcn{J_{4}^z}{\vec{z}_k^\mathfrak{M}-\vec{z}_*^\mathfrak{M}} \Big]
	\end{split}
\end{align}
based on the recent optimal solution $\vec{z}^\mathfrak{M}_{k-1}$:
%\flushleft{
\begin{enumerate}[leftmargin=*]
	\item 	Initialize with:
			\begin{flalign*}
				& \vec{z}^\mathfrak{M}_{0} = \mat{P}_0^z \mat{W}_2^z \, \vec{z}_*^\mathfrak{M} \, , \quad \mat{P}_0^z = \left[ \mat{W}_2^z + \mat{L}_g^z + \mat{L}_c^z \right]^{-1} \, . &
			\end{flalign*}
	\item 	For $k \in \mathbb{N} \setminus \{ 0 \}  $:
			\begin{flalign*}
				 &\vec{z}^\mathfrak{M}_{k} = \vec{z}^\mathfrak{M}_{k-1} + \mat{P}_k^z \vec{m}_k w_{1,k}^z \left[ z_k - \vec{m}_k^T \vec{z}^\mathfrak{M}_{k-1} \right]\, , \\
				 &\mat{P}_{k}^z  = \mat{P}_{k-1}^z - \frac{ \mat{P}_{k-1}^z \vec{m}_k \vec{m}_k^T {\mat{P}_{k-1}^z}^{\!\!\!\!\!\!\!\!\!T} }{{w_{1,k}^z}^{\!\!\!-1} + \vec{m}_k^T \mat{P}_{k-1}^z \vec{m}_k }\, , &
			\end{flalign*}	
		  	where $z_k$ is the new information (the last element of $\vec{z}_k$, cf.\ Eq.\ (\ref{eq:zk})), $\vec{m}_k$ is an abbreviation for \mbox{$\smallfcn{ \vec{m} }{ \vec{x}^\mathfrak{M}, \vec{y}^\mathfrak{M}, x_k, y_k }$}, and $w_{1,k}$ is the last element of $\mat{W}_{1,k}^z$:
%		  		\mbox{$\mat{W}_{1,k} = 
%		  		\left[\begin{array}{c|c}
%		  		\mat{W}_{1,k-1} & \begin{matrix} 0 \\ \vdots \\ 0 \end{matrix} \\ 
%		  		\hline
%		  		\begin{matrix} 0 & \dotsm & 0 \end{matrix} & w_k 
%		  		\end{array}\right] $}.
		  	\begin{flalign*}
		  		& 	\mat{W}_{1,k}^z = 
		  			\left[\begin{array}{c|c}
  						\mat{W}_{1,k-1}^z & \begin{matrix} 0 \\ \vdots \\ 0 \end{matrix} \\ 
						\hline
						\begin{matrix} 0 & \dotsm & 0 \end{matrix} & w_{1,k}^z 
					\end{array}\right] \, . &
			\end{flalign*}
\end{enumerate}
%}
Clearly, step 2 does not differ from the well-known \textit{Recursive Least Squares} (RLS) algorithm; cf.\ \cite[pp.\ 363 ff.]{ljung1999system}.
Consequently, any known issue and modification of the RLS algorithm that can be found in the literature may apply.
The distinguishing feature is the initialization, step 1, where the time-invariant regularization terms are incorporated.
Within the RLS approach, $\mat{P}_k^z$ is the covariance matrix of the estimation error if \mbox{$w_{1,k}^z=1 \ \forall k$} and the information $z_k$ is a normally distributed, uncorrelated signal.
Although this does not apply here, $\mat{P}_k^z$ provides information about the uncertainty of the current estimate $\vec{z}_k^\mathfrak{M}$.
A low diagonal element indicates a reliable estimate of the corresponding element in $\vec{z}_k^\mathfrak{M}$; i.e., considerable information has already been collected within the vicinity of the corresponding grid point.
We refer to these diagonal elements as \textit{uncertainty levels}.

\subsection{Coupled State and Map Estimation}
\label{sec:CSME}

Thus far, the model $(\vec{f}, \vec{g})$, a state estimator (CUKF), and the RME have been presented.
In this section, a combination is presented leading to a novel real-time parameter and state estimation scheme, referred to as \textit{Coupled State and Map Estimator} (CSME), which is expected to be superior if the map parameters have a distinct operating point dependency. % (formerly expressed by $\vec{x}^\mathfrak{M}$--$\vec{y}^\mathfrak{M}$--$\vec{z}^\mathfrak{M}$--$\dots$).
As the overall performance will be sensitive to some implementation details, we propose a specific algorithm, and provide design suggestions.

To avoid an extensive use of indexes, we denote the entry of vector $\vec{x}$ that corresponds to the physical quantity $z$ with $\vec{x}\{z\}$.
Analogously, $\mat{P}\{z\}$ denotes the diagonal element of matrix $\mat{P}$ that corresponds to $z$.
Furthermore, $\vec{D}\{\mat{P}\}$ represents a column vector containing all diagonal elements---in corresponding order---of matrix $\mat{P}$.
The proposed scheme of the CSME is as follows:

\begin{enumerate}[leftmargin=*]
	
	\setlength{\saveleftmarginA}{\leftmargin}
		
	\item 	Declare the required variables, e.g.,
			\begin{flalign*}
				& \hspace*{-\saveleftmarginA} \text{CUKF:}  && \vec{\hat{x}}_0 \, , \quad \mat{R}_0^x\, ,\quad \mat{R}_0^y\, , & \\
				& \hspace*{-\saveleftmarginA} \text{RME}\ (\varphi\ \&\ \mu):  &&\vec{\text{M}}_{d_2}^{\mathfrak{M}}\, , \quad \vec{\Psi}_{p}^{\mathfrak{M}}\, , &\\
				& \hspace*{-\saveleftmarginA} \text{RME}\ (\varphi): &&\vec{\varphi}_*^\mathfrak{M}\, ,  \quad \mat{W}_{2}^\varphi = w_2^\varphi \cdot \mat{I}_{N_\mathfrak{M}}\, , &\\
				&					&& \mat{W}_{g_x}^\varphi = w_g^\varphi\cdot \mat{I}_{N_y}\, , \quad \mat{W}_{g_y}^\varphi = w_g^\varphi\cdot \mat{I}_{N_x}\, , \\
				& 					&& \mat{W}_{c_x}^\varphi = w_{c}^\varphi\cdot \mat{I}_{N_y}\, , \quad \mat{W}_{c_y}^\varphi = w_{c}^\varphi\cdot \mat{I}_{N_x}\, ,&\\
				& \hspace*{-\saveleftmarginA} \text{RME}\ (\mu): &&\vec{\mu}_*^\mathfrak{M}\, ,  \quad \mat{W}_{2}^\mu = w_2^\mu \cdot \mat{I}_{N_\mathfrak{M}}\, , &\\
				&					&& \mat{W}_{g_x}^\mu = w_g^\mu\cdot \mat{I}_{N_y}\, , \quad \mat{W}_{g_y}^\mu = w_g^\mu\cdot \mat{I}_{N_x}\, , \\
				& 					&&  \mat{W}_{c_x}^\mu = w_{c}^\mu\cdot \mat{I}_{N_y}\, , \quad \mat{W}_{c_y}^\mu = w_{c}^\mu\cdot \mat{I}_{N_x}\, .
			\end{flalign*}
			
\end{enumerate}			
The initial state $\vec{\hat{x}}_0$ of the dynamic system refers to all state variables, i.e., normalized temperatures and specific volumes, and variables describing deviations in the work input factor and flow coefficient.
$\mat{I}_N$ denotes an \mbox{$N\times N$} identity matrix.
Remember that $\vec{\text{M}}_{d_2}^{\mathfrak{M}}$ and $\vec{\Psi}_{p}^{\mathfrak{M}}$ are fixed time-invariant grid vectors. % (formerly $\vec{x}^\mathfrak{M}$ and $\vec{y}^\mathfrak{M}$ in Section \ref{sec:map_estimation}).
They have to be preset properly; i.e., they should span the entire range of possible operating points.
Naturally, the number and distribution of the declared grid points determine the flexibility of the performance map and the storage requirement of the routine.
Concerning this trade-off, the aim is to put the maximum compatible number of grid points within the actual domain of possible operating points.
Therefore, we use a rectangular grid with normalized grid points between $\Psi_p=0$ and the expected surge line $\Psi_p=\overline{\Psi}_p(\Mu)$, near which the density of the points increases.
However, if an actual operating point is found to lie outside the preset domain, one could think of applying an extrapolation scheme (the same structure as Eqs\ (\ref{eq:map_phi}) or Eq.\ (\ref{eq:map_mu})) instead of redesigning the interpolation grid.

\noindent Note that every individual compressor stage has its own performance map, and if map variations should be monitored, its own RME calculation steps.
To avoid repetitions, the stage-number superscript Sj is suppressed in this section.
%Just be aware, that every compressor stage require its own performance map vectors and, if map variations shall be monitored, its own RME calculation steps.

\noindent With the scalar weights $w_2^\varphi$ and $w_2^\mu$, the user declares whether to trust the \textit{a priori} performance map \mbox{$\vec{\Psi}_p^\mathfrak{M}$--$\vec{\varphi}_*^\mathfrak{M}$--$\vec{\mu}^\mathfrak{M}_*$--$\vec{\text{M}}_{d_2}^\mathfrak{M}$} (high weights) or not (low weights).
Even in cases where there is no \textit{a priori} knowledge, they have to be declared positive.
For these cases, we set $w_2^\varphi=w_2^\mu=10^{-4}$ and $\vec{\varphi}_*^\mathfrak{M}=\vec{\mu}_*^\mathfrak{M} = \mat{0}_{N_\mathfrak{M} \times 1}  $.
Otherwise, with $w_2=0$, the matrix inverse within the initialization step of the RME may not exist.
Concerning the adjustment of the remaining weights for the presented test case below (see Section \ref{sec:results}), we found a proper balance for \mbox{$w_g^\varphi = w_c^\varphi = {10^{-4}}/{\mat{R}_0^x\{ \Delta \varphi \}}$} and \mbox{$w_g^\mu = w_c^\mu = {10^{-4}}/{\mat{R}_0^x\{ \Delta \mu \}}$}.

\begin{enumerate}[resume, leftmargin=*]

	\setlength{\saveleftmarginA}{\leftmargin}
			
	\item 	Initialize the CUKF and the RME:
			\begin{flalign*}
				& \hspace*{-\saveleftmarginA} \text{CUKF:} &&\vec{\hat{x}}_0\{\Delta \varphi\}=0\, , \quad \vec{\hat{x}}_0\{\Delta \mu\}=0\, , \\
				& 										&& \mat{P}_{x_0}=w_P \cdot \mat{R}_0^x \,  , &\\
				& \hspace*{-\saveleftmarginA} \text{RME}\ (\varphi): && \vec{\varphi}_0^\mathfrak{M} = \mat{P}_0^\varphi \mat{W}_2^\varphi \, \vec{\varphi}_*^\mathfrak{M} \,  , \\
				& 							&& \mat{P}^\varphi_0 = \left[ \mat{W}_2^\varphi + \mat{L}_g^\varphi + \mat{L}_c^\varphi \right]^{-1} , \\
				& \hspace*{-\saveleftmarginA} \text{RME}\ (\mu): && \vec{\mu}_0^\mathfrak{M} = \mat{P}_0^\mu \mat{W}_2^\mu \, \vec{\mu}_*^\mathfrak{M} \,  , \\
				& 							&& \mat{P}^\mu_0 = \left[ \mat{W}_2^\mu + \mat{L}_g^\mu+ \mat{L}_c^\mu \right]^{-1}\,   .
			\end{flalign*}			
			
\end{enumerate}			
Most UT algorithms (cf.\ Section \ref{sec:CUKF}) require $w_P>0$.
Without prior knowledge of $\mat{P}_{x_0}$, $w_P\gg 1$ is a common choice that allows the state estimator to apply large adjustment steps during the initial phase.
Since large adjustments may have a destabilizing effect on the CSME, especially if the prior performance map knowledge is very uncertain ($w_2$ small), we recommend waiting for an initial period before enabling map estimation.
Following this advice, the initial transient behavior of the CUKF, configured via $w_P$, is rather irrelevant in the CSME scheme.

\begin{enumerate}[resume, leftmargin=*]

	\setlength{\saveleftmarginA}{\leftmargin}

	\item 	Initialize the \textit{revised map vectors} (the explanation follows, see step (\ref{item:last})):
			\begin{flalign*}
				& \vec{\widetilde{\varphi}}_0^\mathfrak{M} = \vec{\varphi}_0^\mathfrak{M} \,  , \quad \vec{\widetilde{\mu}}_0^\mathfrak{M} = \vec{\mu}_0^\mathfrak{M} \, .&
			\end{flalign*}		
			
	\item 	\label{item:CSME_k} For $k \in \mathbb{N} \setminus \{ 0 \}  $:
%			\begin{enumerate}[wide, labelwidth=!, labelindent=0pt]
			\begin{enumerate}[leftmargin=*]
			
				\setlength{\saveleftmarginB}{\leftmargin}
				\setlength{\saveleftmargin}{\saveleftmarginA + \saveleftmarginB}
				
				\item 	\label{item:Rxk_b} Adjust time-variant system noise for the CUKF:
						\begin{flalign*}
							 &\mat{R}_{k}^x\{ \Delta \varphi \} = w_{R,{k}} \cdot \mat{R}_0^x\{ \Delta \varphi \}\, , &\\
							 &\mat{R}_{k}^x\{ \Delta \mu \} = w_{R,{k}} \cdot \mat{R}_0^x\{ \Delta \mu \}\, , &\\
							 &\mat{R}_{k}^x\{ \Tr_f^\text{S} \} = w_\Tr \cdot  \left(\vec{f}_{k-1}\{ \Tr_f^\text{S} \}\right)^2 \, , &\\
							 &\text{where} \quad \vec{f}_{k-1} = \smallfcn{\vec{f}}{\vec{\hat{x}}_{k-1},\,\vec{u}_{k-1},\,\vec{\theta},\, t_{k-1}}\, ,&\\
							 &\text{and}  \quad  \vec{\theta}\{ \varphi_i^\mathfrak{M} \} = \widetilde{\varphi}_{i,k-1}^\mathfrak{M}  \, , \quad \vec{\theta}\{ \mu_i^\mathfrak{M} \} = \widetilde{\mu}_{i,k-1}^\mathfrak{M}  \, .&
						\end{flalign*}

				\begin{itemize}[label={},leftmargin=-\saveleftmargin]
					
					\item 	The latter line should clarify that the \textit{revised map vectors} are applied within the compressor stage model.
					As a reminder, $\vec{f}_{k-1}\{ \Tr_f^\text{S} \}$ was introduced in Section \ref{sec:stagemap}, Eq.\ (\ref{eq:dTfS_dt}), as an artificial model equation with the purpose of determining the converged compressor stage's discharge temperature $\Tr_f^\text{S}$ within a small time interval.
					If $\vec{f}_{k-1}\{ \Tr_f^\text{S} \}=0$, then $\Tr_f^\text{S}$ is in a converged state.
					In contrast, if $(\vec{f}_{k-1}\{ \Tr_f^\text{S} \})^2$ is large, then $\Tr_f^\text{S}$ is far from the converged state.
					Enlarging the corresponding model equation uncertainty $\mat{R}_{k}^x\{ \Tr_f^\text{S} \}$ in the latter situation, enables the state estimator to apply large adjustment steps of $\vec{\hat{x}}_k\{ \Tr_f^\text{S} \}$, thus, increasing the speed of convergence.
					We apply $w_\Tr=10^{-2}$.
					
					Further, it is advisable to inform the state estimator whether the current operating point lies within a certain known region (low \textit{uncertainty level}) of the performance map.
					If not, the estimator should be allowed to apply larger deviations from the nominal performance map ($\vec{\hat{x}}_k\{ \Delta \varphi \}$, $\vec{\hat{x}}_k\{ \Delta \mu \}$).
					For instance,
					\begin{flalign*}
%						& \hspace*{\saveleftmargin}	w_{R,k-1}^\varphi = \vec{m}_{k-1}^T \cdot \frac{ \vec{D}\{ \mat{P}_{k-1}^\varphi \}  }{ \min \,\vec{D}\{ \mat{P}_{k-1}^\varphi \} }\, , &\\
						& \hspace*{\saveleftmargin}	w_{R,k} = \vec{m}_{k-1}^T \cdot \frac{ \vec{D}\{ \mat{P}_{k-1}^\mu \}  }{ \min \,\vec{D}\{ \mat{P}_{k-1}^\mu \} } 	&
					\end{flalign*}
					serves this purpose, where $\vec{m}_{k-1}$ is an abbreviation for \mbox{$ \smallfcn{\vec{m}}{ \vec{\text{M}}_{d_2}^{\mathfrak{M}}, \vec{\Psi}_{p}^{\mathfrak{M}}, \widehat{\text{M}}_{{d_2},k-1}, \widehat{\Psi}_{p,k-1} }$},	which contains the interpolation coefficients depending on the location of the recent operating point %\footnote
%					{\label{fn:hat}
%						Superscript  $\ \widehat{~}\ $  denotes the consistent calculation according to the state estimate $\vec{\hat{x}}_{k-1}$ or $\vec{\hat{x}}_{k}$, respectively.
%					}
					within the performance map, which is assumed to lie within the vicinity of the current operating point.
					The hat symbol $\ \widehat{~}\ $  denotes the consistent calculation according to the state estimate $\vec{\hat{x}}$; i.e., respective entries from $\vec{\hat{x}}$ are used to calculate the hat marked values according to the presented formulae.
					Considering that $\vec{D}\{ \mat{P}_{k-1}^\mu \}$ is consistently ordered to the interpolation grid, \mbox{$\vec{m}_{k-1}^T \cdot \vec{D}\{ \mat{P}_{k-1}^\mu \}$} gives the interpolated \textit{uncertainty level} of the recent operating point.
					If the applied interpolation scheme for calculating the interpolation coefficients in $\vec{m}_{k-1}^T$ is comonotone (monotone between neighbored grid points), \mbox{$w_{R,k}\ge 1$} is fulfilled within the entire grid domain (no extrapolation).
					In this case, one could define a \textit{Local Information Level}
					\begin{align}
						 \text{LIL}_{k} :=  \frac{1}{\sqrt{w_{R,k}}}\, , \quad \SI{0}{\percent} < \text{LIL}_k \le \SI{100}{\percent}
					\end{align}
					serving as a meaningful monitoring indicator that correlates with the amount of information collected within the vicinity of the current operating point (``amount of confidence'' in the local estimation).

				\end{itemize}

				\item 	Update the state estimate and the covariance matrix considering the current measurements $\vec{y}_k$ by applying the proposed CUKF scheme:
						\begin{flalign*}
							& \left( \vec{\hat{x}}_k,\, \mat{P}_{x_k} \right) = \fcn{\text{CUKF}}{ \vec{\hat{x}}_{k-1},\, \mat{P}_{x_{k-1}},\, \vec{y}_k, \, \mat{R}^x_{k}, \, \mat{R}^y_k }\, .&
						\end{flalign*}

				\begin{itemize}[label={},leftmargin=-\saveleftmargin]
					
					\item 	The incorporation of the model $(\vec{f}, \vec{g})$ into the CUKF scheme is quite clear (cf.\ Section \ref{sec:CUKF}):
							\begin{flalign*}
								& \hspace*{\saveleftmargin}	\fcn{\vec{F}}{\xkm,\, \vec{u}_{k-1},\, \vec{\theta},\, k} = \xkm \\& \hspace*{4\saveleftmargin}+ \int\limits_{t_{k-1}}^{t_k}  \fcn{\vec{f}}{\vec{x},\,\vec{u}^*,\,\vec{\theta},\, t}\, \diff{t} \, , &\\
								& \hspace*{\saveleftmargin}	\fcn{\vec{G}}{\xk,\, \vec{u}_{k-1},\, \vec{\theta},\, k} = \fcn{\vec{g}}{\vec{x},\,\vec{u},\,\vec{\theta},\, t_k}\, .&
							\end{flalign*}
							From the user's point of view, note that we end up using a simple forward Euler method for numerical integration that provides acceptable performance in terms of stability, accuracy, and computational speed, at least with step sizes $\approx \SI{5e-2}{\second}$ for volumes $\ge \SI{e-2}{\cubic\meter}$, as chosen here.
							Further, since entries in the input vector $\vec{u}$ that comprises the individual shaft speeds and the last stage's discharge pressure may arise from time-sampled measurements (cf.\ Eq.\ (\ref{eq:u})), it is not advisable to ignore their changes throughout the prediction horizon, especially in cases of highly transient operation or large $t_{k}-t_{k-1}$.
							Therefore,
							\begin{flalign*}
								& \hspace*{\saveleftmargin}	\fcn{\vec{u}^*\!\!}{t} = \left[ 1 - \frac{t - t_{k-1}}{t_k - t_{k-1}} \right]\cdot \vec{u}_{k-1} + \frac{t - t_{k-1}}{t_k - t_{k-1}}\cdot \vec{u}_k&
							\end{flalign*}
							is embedded in the numerical integration scheme.

				\end{itemize}

				\item 	If the initial transient phase of the state estimator is concluded, continue with step (\ref{item:next}); otherwise, skip steps (\ref{item:next})--(\ref{item:last}).
				
				\begin{itemize}[label={},leftmargin=-\saveleftmargin]
					
					\item	\label{item:startRLS} 	For a proper indication, condition 
							\begin{flalign*}
								& \hspace*{\saveleftmargin}  \left\lVert	\left[\vec{D}\{\mat{P}_{x_k}\} - \vec{D}\{\mat{P}_{x_{k-1}}\}\right] \oslash \vec{D}\{\mat{P}_{x_{k-1}}\} \right\rVert_\infty \\
								& \hspace*{5.5\saveleftmargin} < \text{``threshold''}  &
							\end{flalign*}  
							may be checked, where the operator $\oslash$ denotes the Hadamard division (element-wise division).
							
				\end{itemize}

				\item 	\label{item:next} Concerning the weighted map estimation error (cf.\ Eq.\ (\ref{eq:e_map})), adjust the time-variant weights for the RME:
						\begin{flalign*}				
							& w_{1,k}^\varphi = \frac{1}{ \mat{P}_{x_k}\{ \Delta \varphi \} } \, , \quad w_{1,k}^\mu = \frac{1}{ \mat{P}_{x_k}\{ \Delta \mu \} }\, . &
						\end{flalign*}
						
				\begin{itemize}[label={},leftmargin=-\saveleftmargin]
	
				\item 	In a standard \textit{Least Squares} approach without regularization terms, $\mat{W}_{1,k}^z = ({\mat{C}^z_k})^{-1} $ gives the optimal (minimum covariance) estimate of $\vec{z}^\mathfrak{M}_k$ if $\mat{C}^z_k$ is the true covariance matrix of the collected information $\vec{z}_k$ that arises from a stochastic, uncorrelated process.
				We already stated that such a premise does not apply here.
				The current information to be considered will arise from the CUKF estimate $\vec{\hat{x}}_k\{z\}$, which is treated as the expected mean of an unspecified distribution with an expected variance $\mat{P}_{x_k}\{ z \} $.
				However, the proposed weighting $w_{1,k}^\varphi$ and $w_{1,k}^\mu$ clearly indicates the underlying intention.
				
				\end{itemize}						
								
				\item 	\label{item:RMEupdate} Incorporate the updated state estimates $\vec{\hat{x}}_k \{ \Delta \varphi \}$ and $\vec{\hat{x}}_k \{ \Delta \mu \}$ into the performance map applying the recursive step of the RME scheme:%, each for $\vec{\varphi}^\mathfrak{M}$ and $\vec{\mu}^\mathfrak{M}$:
%						\begin{flalign*}
%							& \hspace*{-\saveleftmargin} \text{RME}\ (\varphi): &&  \left( \vec{\varphi}_k^\mathfrak{M} ,\, \mat{P}_k^\varphi \right) = \fcn{\text{RME}}{ \vec{\varphi}_{k-1}^\mathfrak{M}, \, \mat{P}_{k-1}^\varphi , \, \varphi_k , \, \vec{m}_k, \, w_{1,k}^\varphi } , &\\
%							& \hspace*{-\saveleftmargin} \text{RME}\ (\mu): &&  \left( \vec{\mu}_k^\mathfrak{M} ,\, \mat{P}_k^\mu \right) = \fcn{\text{RME}}{ \vec{\mu}_{k-1}^\mathfrak{M}, \, \mat{P}_{k-1}^\mu , \, \mu_k , \, \vec{m}_k, \, w_{1,k}^\mu }\, , &
%						\end{flalign*}
						\begin{flalign*}
							& \hspace*{-0\saveleftmargin} \text{RME}\ (\varphi): &  \vec{\varphi}^\mathfrak{M}_{k} &= \vec{\varphi}^\mathfrak{M}_{k-1} + \mat{P}_k^\varphi \vec{m}_k w_{1,k}^\varphi \Big[ \widehat{\varphi}_k &\\
							& & &\hspace*{0.3cm}- \vec{m}_k^T \vec{\varphi}^\mathfrak{M}_{k-1} \Big] , &\\
							& & \mat{P}_{k}^\varphi  &= \mat{P}_{k-1}^\varphi - \frac{ \mat{P}_{k-1}^\varphi \vec{m}_k \vec{m}_k^T {\mat{P}_{k-1}^\varphi}^{\!\!\!\!\!\!\!\!\!T} }{{w_{1,k}^\varphi}^{\!\!\!-1} + \vec{m}_k^T \mat{P}_{k-1}^\varphi \vec{m}_k }\, , & \\
							& \hspace*{-0\saveleftmargin} \text{RME}\ (\mu): &  \vec{\mu}^\mathfrak{M}_{k} &= \vec{\mu}^\mathfrak{M}_{k-1} + \mat{P}_k^\mu \vec{m}_k w_{1,k}^\mu \Big[ \widehat{\mu}_k \\
							& & &\hspace*{0.3cm}- \vec{m}_k^T \vec{\mu}^\mathfrak{M}_{k-1} \Big] , \\
							& & \mat{P}_{k}^\mu  &= \mat{P}_{k-1}^\mu - \frac{ \mat{P}_{k-1}^\mu \vec{m}_k \vec{m}_k^T {\mat{P}_{k-1}^\mu}^{\!\!\!\!\!\!\!\!\!T} }{{w_{1,k}^\mu}^{\!\!\!-1} + \vec{m}_k^T \mat{P}_{k-1}^\mu \vec{m}_k }\, , &
						\end{flalign*}
						
				\begin{itemize}[label={},leftmargin=-\saveleftmargin]
					
					\item 	where the ``new information'' to be considered arises from 
							\begin{flalign*}
								& \hspace*{\saveleftmargin} \text{RME}\ (\varphi): & \widehat{\varphi}_k = \vec{m}_k^T \cdot \vec{\widetilde{\varphi}}_{k-1}^\mathfrak{M} + \vec{\hat{x}}_k \{ \Delta \varphi \}\, , \hspace*{8cm} &\\
								& \hspace*{\saveleftmargin} \text{RME}\ (\mu): & \widehat{\mu}_k = \vec{m}_k^T \cdot\vec{\widetilde{\mu}}_{k-1}^\mathfrak{M} + \vec{\hat{x}}_k \{ \Delta \mu \}\, , \hspace*{8cm} &
							\end{flalign*}
					and	$\vec{m}_{k}$ is an abbreviation for \mbox{$ \smallfcn{\vec{m}}{ \vec{\text{M}}_{d_2}^{\mathfrak{M}}, \vec{\Psi}_{p}^{\mathfrak{M}}, \widehat{\text{M}}_{{d_2},k}, \widehat{\Psi}_{p,k} }$}.  %\footnotemark[\ref{fn:hat}]
					Note that, for instance, \mbox{ $\vec{m}_k^T \cdot \vec{\widetilde{\varphi}}_{k-1}^\mathfrak{M}$} is embedded in the model in place of $\Sj{\varphi}$ from Eq.\ (\ref{eq:mS1}), and $\vec{\hat{x}}_k \{ \Delta \varphi \}$ represents $\Sj{\Delta \varphi}$ in this context; i.e., the CUKF calculates the displacement in relation to the former \textit{revised map}.
					
				\end{itemize}	
				
				\item 	\label{item:last} Update the \textit{revised map vectors} with the approach described below:
						\begin{flalign*}
							& \vec{\widetilde{\varphi}}_k^\mathfrak{M} =  \Az \left[ \Az^T \mat{W}_k^{\widetilde{\varphi}} \Az \right]^{-1} \Az^T \mat{W}_k^{\widetilde{\varphi}} \left[ \vec{\varphi}_k^\mathfrak{M} - \bv \right] + \bv\, ,&\\
							& \vec{\widetilde{\mu}}_k^\mathfrak{M} =  \Az \left[ \Az^T \mat{W}_k^{\widetilde{\mu}} \Az \right]^{-1} \Az^T \mat{W}_k^{\widetilde{\mu}} \left[ \vec{\mu}_k^\mathfrak{M} - \bm \right] + \bm \, ;& 
						\end{flalign*}
						and reset the \textit{a priori} performance map deviations of the next iteration afterward:
						\begin{flalign*}
							& \vec{\hat{x}}_{k}\{\Delta \varphi\}=0\, , \quad \vec{\hat{x}}_{k}\{\Delta \mu\}=0\, . &
						\end{flalign*}						
						
				\begin{itemize}[label={},leftmargin=-\saveleftmargin]		
					
					\item 	Before the newly introduced variables are declared, the conceptual idea behind the \textit{revised map vectors} and step (\ref{item:last}) needs clarification.
					A sketch of this concept is shown in Fig.\ \ref{fig:updateMap} for the simplified situation $ \widehat{\text{M}}_{d_2,{k-1}} = \widehat{\text{M}}_{d_2,{k}} = \text{M}_{d_2,i}^\mathfrak{M} $; i.e., the dimension along machine Mach number variation becomes neglectable, yielding a scalar interpolation approach along $\Psi_{p}$ only.
					\begin{figure}
						\centering
						\includegraphics[width=1\columnwidth]{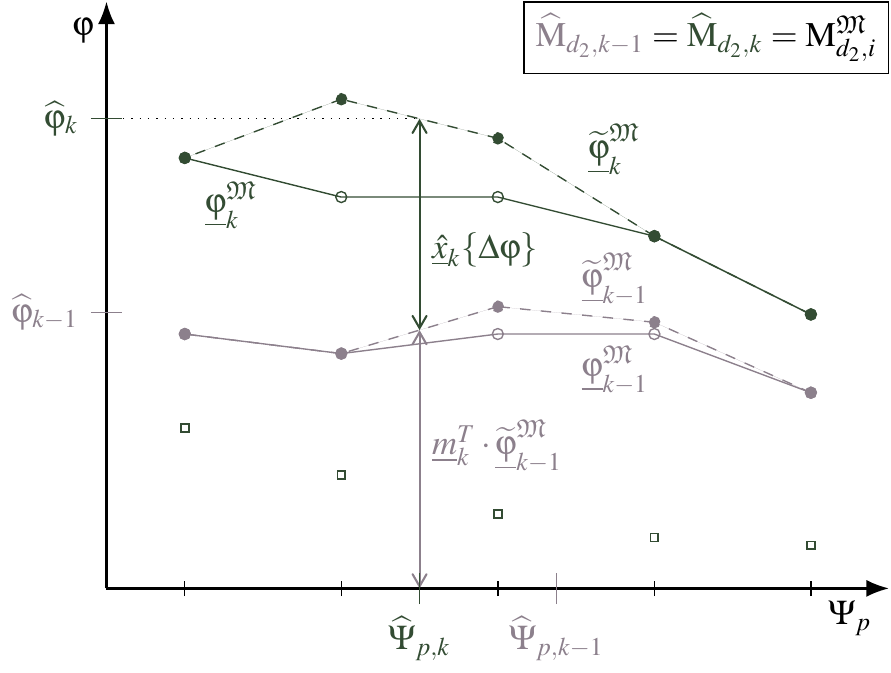}
						\caption[Revised Map]{Update of revised map vector $\widetilde{\vec{\varphi}}^\mathfrak{M}$ applying a scalar C\textsuperscript{0} continuous linear interpolation approach (dependence on $\Mu$ neglected); symbols \raisebox{1pt}{ \tikz{ \draw[line width=0.5pt,scale=0.3] (0em,0em) rectangle ++(1em,1em); } } denote \textit{uncertainty levels} at grid points  }
						\label{fig:updateMap}
					\end{figure}
					For the current time stamp $k$, the CUKF estimates a deviation $\vec{\hat{x}}_k\{\Delta \varphi\}$ to the former \textit{revised map}, the $\varphi$-coordinates of which are stored in $\vec{\widetilde{\varphi}}_{k-1}^\mathfrak{M}$.
					In step (\ref{item:RMEupdate}) this deviation is incorporated into the \textit{actual map}, the $\varphi$-coordinates of which are stored in $\vec{\varphi}_k^\mathfrak{M}$, by recursively solving the optimization problem, as described in Section \ref{sec:map_estimation}.
					As is the situation in Fig.\ \ref{fig:updateMap}, the estimated flow coefficient $\widehat{\varphi}_k$ is unlikely to have no bias to the actual map (normally, an interpolated surface; here, an interpolated line); i.e., \mbox{ $\vec{m}_k^T \cdot \vec{\varphi}_{k}^\mathfrak{M} \ne \widehat{\varphi}_k$}.
					In other words, the actual map is not consistent with the estimated state.
					This is expected, since the CUKF presumes the map to be time-invariant. % and a time-variant deviation ($\mat{R}_k^x\{ \Delta \varphi \} > 0$).
					Instead of advancing the model states by $\vec{\varphi}_k$, etc., and applying a joint estimation within the CUKF scheme, which would be in accordance with the approach in \cite{hoeckerdal2011ekf} and raise a massive increase in the computational burden in the present context,
%					\footnote
%					{
%						Such an approach raises observability problems and a massive increase in the computational burden.
%					}
					we derived the \textit{Coupled State and Map Estimator} and provide the revised map vectors for this purpose.
					These vectors, $\vec{\widetilde{\varphi}}_{k}^\mathfrak{M}$ and $\vec{\widetilde{\mu}}_{k}^\mathfrak{M}$, are consistent to the CUKF estimate, i.e., 
					\begin{align}
						\vec{m}_k^T \cdot \vec{\widetilde{\varphi}}_{k}^\mathfrak{M} = \widehat{\varphi}_k\, , \quad \vec{m}_k^T \cdot \vec{\widetilde{\mu}}_{k}^\mathfrak{M} = \widehat{\mu}_k \, ,\label{eq:NB}
					\end{align}
				 	which is assured via step (\ref{item:last}).
					Metaphorically speaking, the estimated deviation is preserved within the revised map.
					Therefore, the respective CUKF states must be reinitialized for the next iteration, to be consistent with the revised situation itself.
					Consequently, the revised map and the CUKF estimates are capable of tracking spontaneous and wide map deviations quickly, even if the actual map, calculated within the RME scheme, is in a nearly converged state, where adjustments to changed circumstances are typically sluggish.\footnote
					{
						Forgetting factors may be incorporated in the RME to keep the actual map flexible.
						Within the presented context, global forgetting does not make sense, and---as far as our experience goes---local forgetting is hard to adjust properly.
						Alternatively, the CSME might be restarted frequently, e.g., with $\vec{\varphi}_*^\mathfrak{M} = \vec{\varphi}_{k-1}^\mathfrak{M}$, etc.
					}
					The diverging response times of actual and revised map vectors may be exploited for fault detection.
					As an example for $\mu$, if condition 
%					\begin{align}
%						\text{i)} \ \left\lVert	\vec{\mu}_j^\mathfrak{M} - \vec{\widetilde{\mu}}_j^\mathfrak{M}  \right\rVert_\infty > w_{f}\, \sqrt{ \mat{P}_{x_j}\{ \Delta \mu \} }\, , \quad \text{ii)} \ \text{LIL}_j > \epsilon
%					\end{align}
					\begin{align}
						\left\lVert	\vec{\mu}_j^\mathfrak{M} - \vec{\widetilde{\mu}}_j^\mathfrak{M}  \right\rVert_\infty > w_{f}\, \sqrt{ \mat{P}_{x_j}\{ \Delta \mu \} } \label{eq:FD}
					\end{align}
					is fulfilled for $(N_f+1)$ consecutive time stamps \mbox{($j = \{ k-N_f,\, \dots,\, k-1,\, k \}$)}, a drastic change in behavior, i.e., a fault at time stamp $k-N_f$, of the corresponding compressor stage is plausible.
					
					The proposed calculation of $\vec{\widetilde{\varphi}}_{k}^\mathfrak{M}$ or $\vec{\widetilde{\mu}}_{k}^\mathfrak{M}$ minimizes the weighted sum of squared errors between $\vec{\varphi}_{k}^\mathfrak{M}$ and $\vec{\widetilde{\varphi}}_{k}^\mathfrak{M}$ or between $\vec{\mu}_{k}^\mathfrak{M}$ and $\vec{\widetilde{\mu}}_{k}^\mathfrak{M}$, respectively, subjected to (\ref{eq:NB}).
					We found it reasonable to keep revised points with a low \textit{uncertainty level} close to the actual map (cf.\ Fig.\ \ref{fig:updateMap}).
%					allow larger deviations for grid points with a higher \textit{uncertainty level}.
					Therefore, weighting matrices $\mat{W}_k^{\widetilde{z}}$ are constructed as diagonal matrices that fulfill $\vec{D}\{ {\mat{W}_k^{\widetilde{z}}}^{-1} \} =	\vec{D}\{ {\mat{P}_k^{z}} \}$.
%					\begin{align*}
%						\vec{D}\{ {\mat{W}_k^{\widetilde{\varphi}}}^{-1} \} =	\vec{D}\{ {\mat{P}_k^{\varphi}} \}\, .
%					\end{align*}

					$\Az$, $\bv$, and $\bm$ are as follows:
							\begin{flalign*}
	%							&  \kbordermatrix
	%								{		& 1			& 2 & 			& j-2 	& j-1 	& j & 			& N_\mathfrak{M}-1 	\\
	%									1	& 1			& 0	& \dotsm 	& 0		& 0   	& 0 & \dotsm 	& 0					\\
	%									2	& 0			& 1	& \dotsm 	& 0		& 0   	& 0 & \dotsm 	& 0					\\
	%										& \vdots	& 	& \ddots 	&  		& 0   	&  	&  			& \vdots			\\
	%								  j-2 	& 0 		& 0 & 			& 1		& 0 	& 0 & 			& 0					\\
	%								  j-1 	& 0 		& 0 & 			& 0		& 1 	& 0 & 			& 0
	%								}		&
					  			 \hspace*{\saveleftmargin}	\Az &= 
					  			\begingroup
						  			\renewcommand*{\arraystretch}{1.5}
									\left[\begin{array}{ccc}
										\mat{I}_{{j_k}-1} 	 & \vrule &   \mat{0}_{({j_k}-1)\times(N_\mathfrak{M}-{j_k})} \\
										\hline \\[-1.5\normalbaselineskip]
										\multicolumn{3}{c}{ \smash{\raisebox{.1\normalbaselineskip}{$-m_{k,{j_k}}^{-1}\cdot \vec{m}_{{j_k},k}^T$}} } \\
										\hline \\[-1.5\normalbaselineskip]
										\mat{0}_{(N_\mathfrak{M}-{j_k})\times({j_k}-1)}  &\vrule&  \mat{I}_{N_\mathfrak{M}-{j_k}}
									\end{array}\right]
								\endgroup \, , & \\ 
								\hspace*{\saveleftmargin} \bv &=
								\begingroup
									\renewcommand*{\arraystretch}{1.5}
									\left[\begin{array}{c}
										\mat{0}_{({j_k}-1)\times 1} \\
										\hline \\[-1.5\normalbaselineskip]
										\smash{\raisebox{.1\normalbaselineskip}{ $\widehat{\varphi}_k \cdot m_{k,{j_k}}^{-1}$ }} 	\\
										\hline \\[-1.5\normalbaselineskip]
										\mat{0}_{(N_\mathfrak{M}-{j_k})\times 1}
									\end{array}\right]
								\endgroup \, , \quad \bm =
								\begingroup
								\renewcommand*{\arraystretch}{1.5}
								\left[\begin{array}{c}
								\mat{0}_{({j_k}-1)\times 1} \\
								\hline \\[-1.5\normalbaselineskip]
								\smash{\raisebox{.1\normalbaselineskip}{ $\widehat{\mu}_k \cdot m_{k,{j_k}}^{-1}$ }} 	\\
								\hline \\[-1.5\normalbaselineskip]
								\mat{0}_{(N_\mathfrak{M}-{j_k})\times 1}
								\end{array}\right]
								\endgroup \, .&
							\end{flalign*}
							$m_{k,{j_k}}$ is the $j_k$th element of $\vec{m}_k$, and $\vec{m}_{{j_k},k} \in \mathbb{R}^{N_\mathfrak{M}-1}$ is a subvector of $\vec{m}_{k}$, constructed by removing $m_{k,{j_k}}$, and $j_k$ is an arbitrary index that fulfills $m_{k,{j_k}}\ne0$.
							Be aware that---applying a C\textsuperscript{0} continuous interpolation method---the matrix inversion in step (\ref{item:last}) can be reformulated; thus, the actual matrix to be inverted is of dimension $3\times 3$.

				\end{itemize}
				
	\end{enumerate}
	
\end{enumerate}

%%%%%%%%%%%%%%%%%%%%%%%%%%%%%%%%%%%%%%%%%%%%%%%%%%%%%%%%%%%%%%%%%%%%%%
\section{Results}
\label{sec:results}

\subsection{Test Case}
\label{sec:setup}

For validation purposes, a reference process with known performance maps and state trajectories is mandatory, and distinct real-gas behavior is desired to emphasize the scope of this research.
Therefore, a numerical, i.e., simulative experiment (SE) of a two-stage supercritical \COO\ (\mbox{$T_c=\SI{31.13}{\celsius}$}, \mbox{$p_c=\SI{73.77}{\bar}$}) compressor acts as reference and ``measurement'' generator.
The ``measurements'' are sampled at a rate of \SI{1}{\hertz}.
At a reasonable effort, several modifications have been implemented compared to the model that is applied within the monitoring scheme (cf.\ Section \ref{sec:model}), to draw a somewhat more realistic situation for the CSME:
\begin{itemize}[label=$\bullet$, leftmargin=*]
	
	\item In contrast to the approximative but versatile real-gas model that is applied within the CSME (cf.\ Section \ref{sec:realgas}), the \COO-specific model presented in \cite{span1996new}, which matches the real-gas behavior of \COO\ closely, is embedded within the SE.
	
	\item The assumption of well-mixed volumes within the connecting pipes is discarded, and the delay of the temperature information due to (1D) transportation of the mass inside the pipes is considered in the SE.
	
	\item For the mapping function $\mathfrak{M}$, a C\textsuperscript{1} continuous (piecewise cubic) interpolation method is applied based on 350 grid points vs. C\textsuperscript{0} continuous (bilinear) interpolation with \mbox{$N_\mathfrak{M}=140$} grid points in the CSME.
	
	\item For every time stamp, the converged discharge temperature is calculated vs. the artificial state approach according to Eq.\ (\ref{eq:dTfS_dt}) of Section \ref{sec:stagemap}, in the CSME.
	
\end{itemize}

%Additionally, two issues concerning the compressor stage model differ (vsee the model applied within the CSME):
%; ii) 

\begin{figure*}
	\centering
	\includegraphics[width=\textwidth]{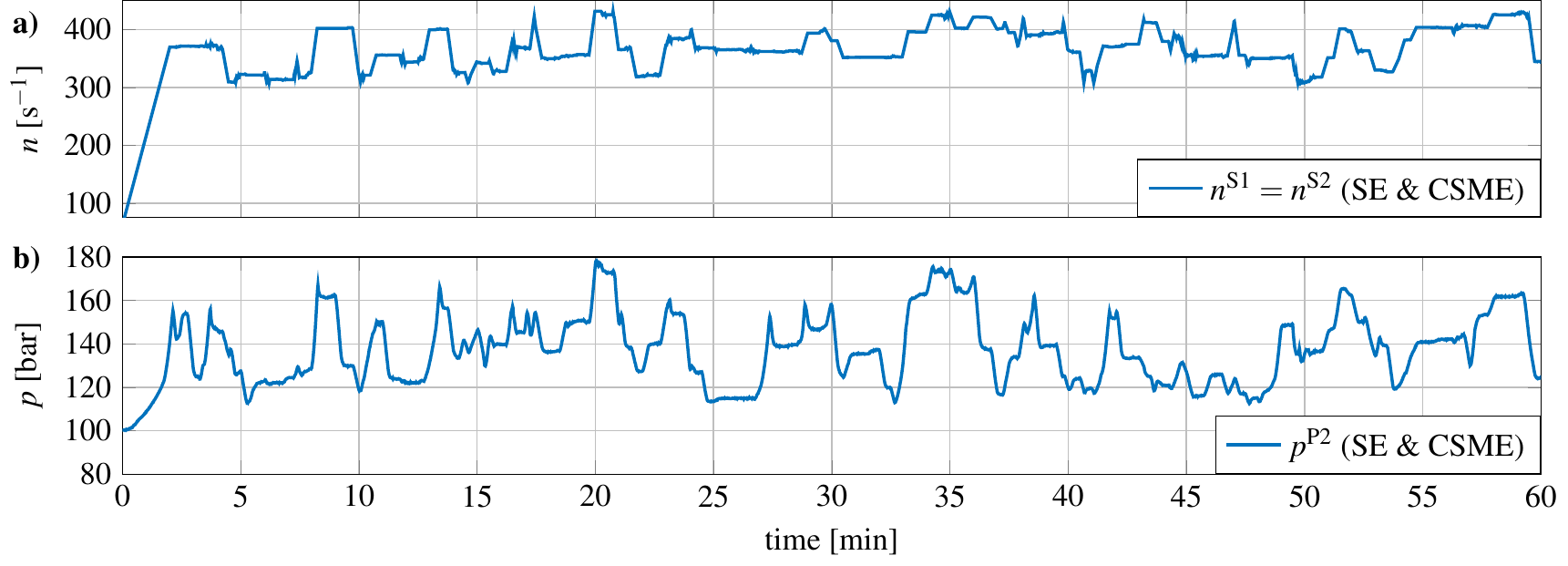}
	\caption[Simulation inputs]{Simulation inputs;\\ \ a) identical compressor-shaft speed for both stages S1 and S2; b) discharge pressure $\Pc{p}$ of discharge pipe P2.}
	\label{fig:SE_Input}
\end{figure*}

The preset simulation inputs, the compressor-shaft speed and the discharge pressure, can be found in Fig.\ \ref{fig:SE_Input}.
We do not claim to have designed a realistic operating scenario.
The intention was to run the machine across varying operating points, as in a highly flexible operation, connected via transients with a supposedly realistic, non-stepwise shape.
The suction conditions of the SE are fixed at \mbox{$\Pa{p}=\SI{100}{\bar}$}, \mbox{$\Pa{T}=\SI{80}{\celsius}$}.
All calculated values that are treated as ``measurements'' for the CSME are affected by artificial, normally distributed noise with a standard deviation of $\SI{0.1}{\bar}$, $\SI{0.1}{\kelvin}$, or $\SI{0.1}{\kg\per\second}$, respectively.
The time stamp index $k$ is suppressed in this section, since the physical time, e.g., measured in \textit{minutes}, seems more natural.
Such a dependence is obvious from the following figures anyway.

\subsection{Process Tracking}
\label{sec:tracking}

\begin{figure*}
	\centering
	\includegraphics[width=\textwidth]{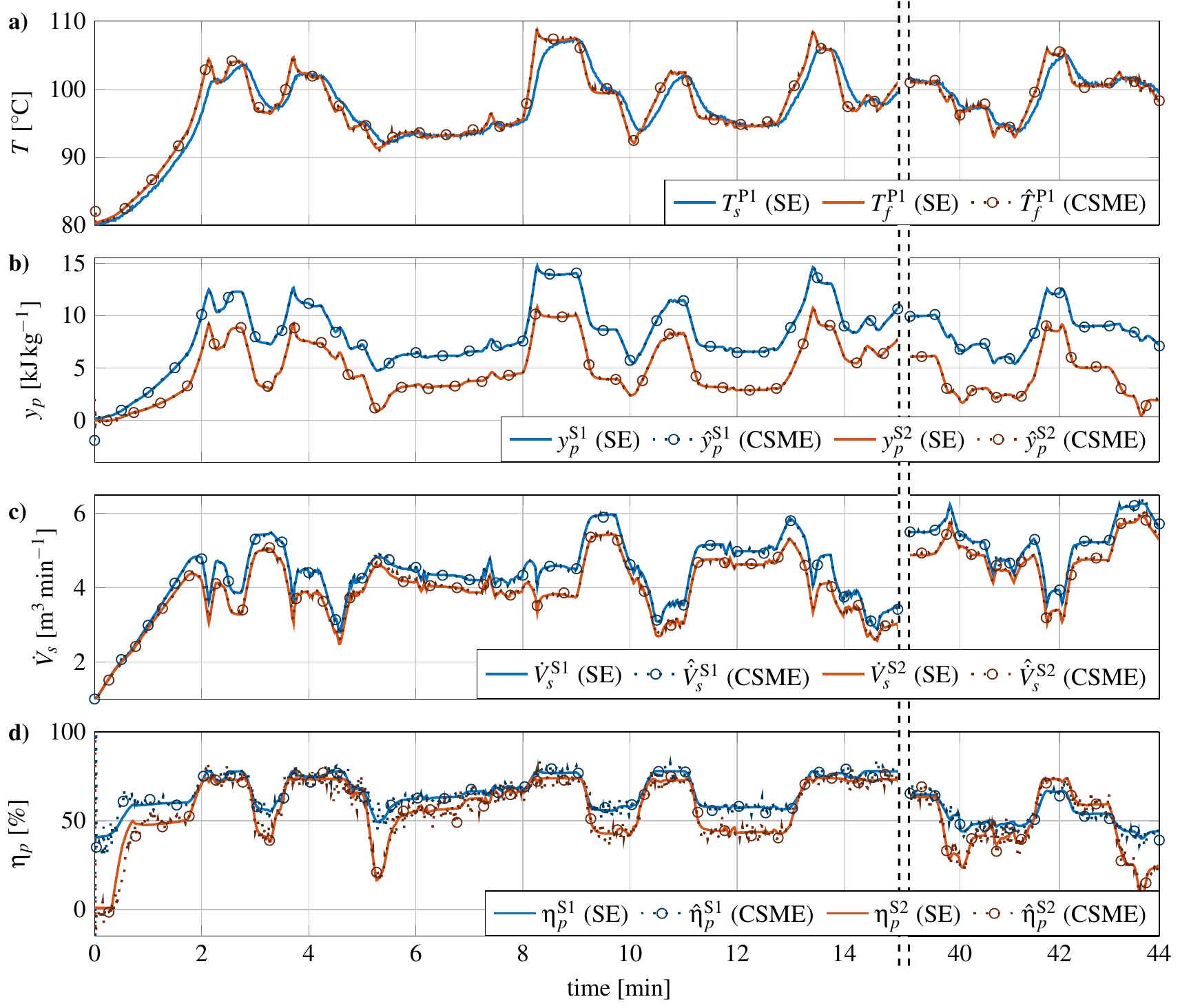}
	\caption[Tracking performance of the CSME]{Tracking performance of the CSME for $t \in \left[ \SI{0}{\minute}; \SI{15}{\minute} \right] \cup \left[ \SI{39}{\minute}; \SI{44}{\minute} \right]$; true trajectories are depicted with solid lines whereas estimated trajectories are depicted with line shape \raisebox{0pt}{\tikz{\draw[-,loosely dotted,thick,scale=0.9](0em,0.5em)--(2.2em,0.5em); \draw[thick,scale=0.9] (1.05em,0.5em) circle (0.3em); }};\\ \ a) intermediate temperatures; b) polytropic heads; c) suction volume flows; d) polytropic efficiencies}
	\label{fig:SE_Track}
\end{figure*}

%Concerning the tracking performance, i.e., the capability of the CSME to track unmeasured (in reality unknown) process states, some exemplary results are shown in Fig.\ \ref{fig:SE_Track}.
Concerning the tracking performance, i.e., the capability of the CSME to track reference process states, we focus on the unmeasured, in reality unknown states.
The CSME is able to track the measurement values as well, while significantly reducing the (artificial) measurement noise.
Some exemplary results are shown in Fig.\ \ref{fig:SE_Track}.
%As can be seen in Fig.\ \ref{fig:SE_Track}c, when comparing the intermediate temperature of the fluid $\Pb{T_f}$ and the corresponding temperature at sensor location $\Pb{T_s}$, both calculated in the SE, the temperature measurements are delayed.
The delay between the fluid's temperature in the intermediate pipe $\Pb{T_f}$ and the corresponding ``measured'' value $\Pb{T_s}$, both calculated in the SE, can be seen in Fig.\ \ref{fig:SE_Track}a for the intermediate temperature. %\footnote
%{
%	The setting is $\Pb{\tau}=\SI{10}{\second}$ (cf.\ Eq.\ (\ref{eq:dTs_dt})) for the SE as well as for the CSME.
%}
In the SE and the CSME, a time constant of $\Pb{\tau}=\SI{10}{\second}$ is assumed for a first-order system; see Section \ref{sec:state_eq}, Eq.\ (\ref{eq:dTs_dt}).
The CSME is capable of tracking the actual temperature of interest $\Pb{T_f}$ (slightly noisy), revealing any temperature peak, which is hidden from the measurements.
For the remaining temperature positions, the tracking performance is very similar; thus, they are omitted here.

Fig.\ \ref{fig:SE_Track}b--d shows the estimation of several compressor characteristics. %, which are not affected by transient behavior as the common steady-state calculation would.
Obviously, the generic real-gas model, presented in Section \ref{sec:realgas}, is sufficiently accurate; otherwise, the estimates would have to be biased from the reference.
None of the values shown in Fig.\ \ref{fig:SE_Track}b--d are declared model states for the CUKF scheme; thus, the values arise from the presented formulae (see Section \ref{sec:stagemap}), embedding the ``direct estimates'' preserved in $\vec{\hat{x}}$.
Since $\vec{\hat{x}}\{ \Delta \mu \}$ enters the denominator of 
$$\hat{\eta}_p = \frac{\widehat{\Psi}_{p}}{2\, \widehat{\mu}} = \frac{\widehat{\Psi}_{p,k}}{2\, \left( \vec{m}_k^T \cdot\vec{\widetilde{\mu}}_{k-1}^\mathfrak{M} + \vec{\hat{x}}_k \{ \Delta \mu \} \right)}$$
(cf.\ Eq.\ (\ref{eq:etap}) and step (\ref{item:RMEupdate}) of Section \ref{sec:CSME}), this value is particularly prone to noise transmission, as can be seen in Fig.\ \ref{fig:SE_Track}d.
Note that the tuning parameters, e.g., $\mat{R}_0^x$, $\mat{R}^y_k$, weights, etc., are manually tuned, and thus, the existence of an alternative parameter set that provides a ``better'' performance is very likely.

\subsection{Fault Indication and Isolation}
\label{sec:FDI}

Although the term \textit{alteration} would be much better suited in the context of this work, the common term \textit{fault} is used consistently.
To investigate the simple fault detection scheme stated in Section \ref{sec:CSME} (cf.\ Eq.\ (\ref{eq:FD})), the reference performance map of the \textit{first compressor stage} is modified within the SE from $t=\SI{40}{\minute}$ on, referred to as a fault event in the present section.
More specifically, the first stage's work input factor, calculated via a reference mapping function, is increased by $0.1$ for $t\ge\SI{40}{\minute}$ in the SE.
A respective, spontaneous efficiency decline can already be found in Fig.\ \ref{fig:SE_Track}d, while it is hard to detect an increase in the temperature measurement in Fig.\ \ref{fig:SE_Track}a during flexible operation.
Altogether, the tracking performance seems unaffected by the fault, which indicates a correct fault isolation.
Generally, fault isolation is the capability to assign a fault to the correct cause.
Here, due to the lack of defining particular failure sets, the term fault isolation is used for the local assignment to a specific plant component, i.e., the correct compressor stage.
Because the proposed fault detection is based on component-specific parameters, fault isolation is a straightforward task.

\begin{figure*}
	\centering
	\includegraphics[width=\textwidth]{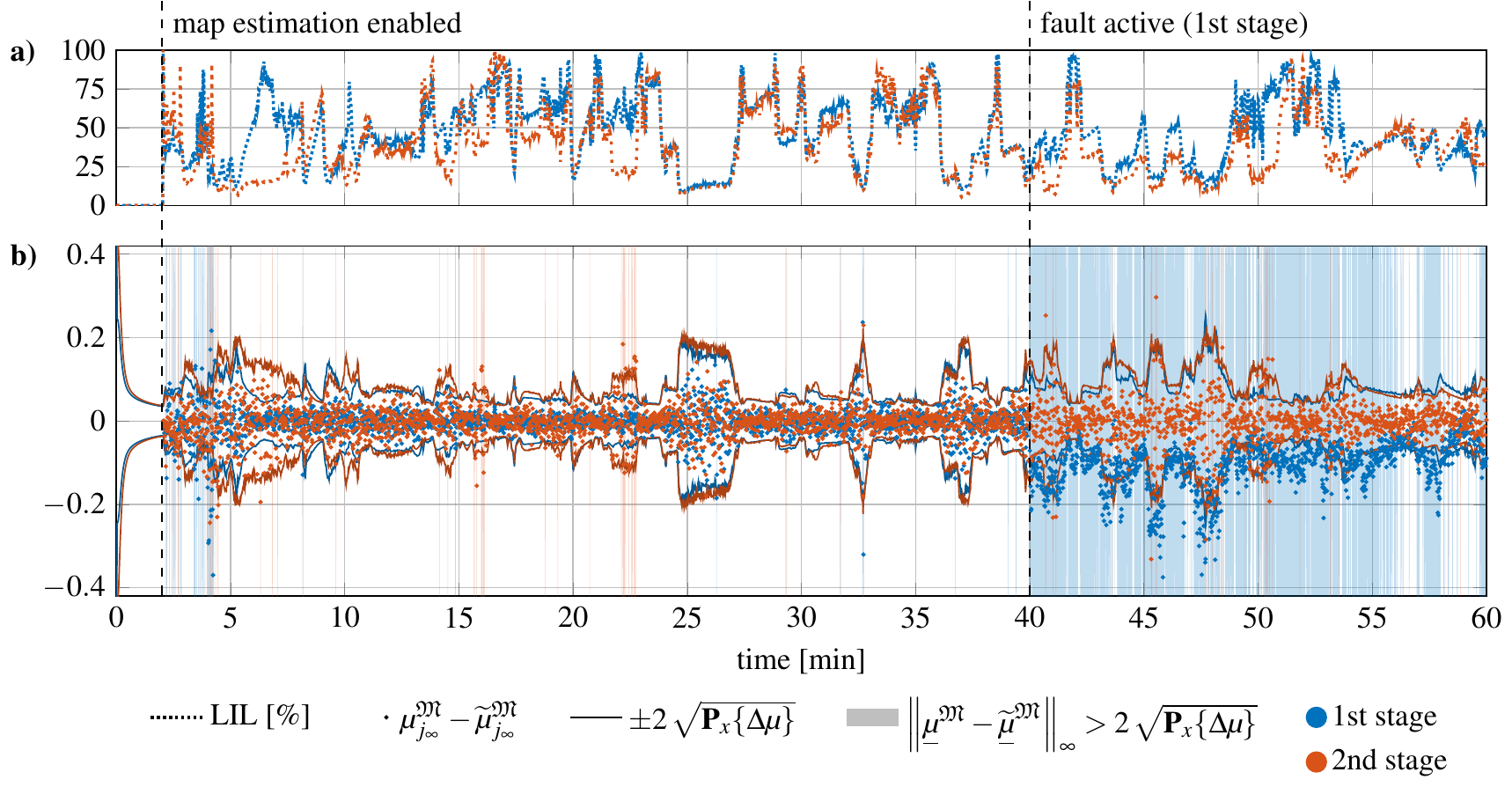}
	\caption[Fault indication and isolation]{Fault indication and isolation; \\ \ index $j_\infty$ is defined via \mbox{$ \left| {\mu}^\mathfrak{M}_{j_\infty} - {\widetilde{\mu}}^\mathfrak{M}_{j_\infty} \right| = \left\lVert	\vec{\mu}^\mathfrak{M} - \vec{\widetilde{\mu}}^\mathfrak{M}  \right\rVert_\infty $}}
	\label{fig:SE_FDI}
\end{figure*}

Situations where the fault condition (\ref{eq:FD}) is fulfilled for the current time stamp (\mbox{$N_f=0$}) setting \mbox{$w_f=2$} are referred to as \textit{fault indication}.
A fault indication is marked by a colored background in Fig.\ \ref{fig:SE_FDI}b.
%As can be anticipated, a proper selection of $N_f$ can eliminate false detections.
The proposed scheme clearly indicates the fault assigned to the first compressor stage.
For $N_f=5$ (six consecutive fault indications), an automated fault detection would have raised the failure flag for ``compressor stage 1'' for the first time \SI{7}{\second} after the actual fault event, while failure flag ``compressor stage 2'' would remain deactivated for the entire experiment. 
Naturally, the frequency of the fault indications diminishes over time, since the actual map estimate $\vec{\mu}^\mathfrak{M}$ is never in a converged state if a failure has occurred.
Consequently, the faulty behavior is gradually incorporated into the actual map estimate.

In Fig.\ \ref{fig:SE_FDI}, the advantage of defining a time-variant threshold ($2\, \sqrt{\mat{P}_{x_k}\{ \Delta \mu \}}$) for fault indication is quite obvious.
Every time the compressor stage runs into an uncertain operating range, i.e., the \textit{Local Information Level} LIL is low (Fig.\ \ref{fig:SE_FDI}a), the amplitude of adaption, recognizable via $\mu_{j_\infty}^\mathfrak{M} - \widetilde{\mu}_{j_\infty}^\mathfrak{M}$ (Fig.\ \ref{fig:SE_FDI}b), increases, which is enhanced by step \mbox{%\ref{item:CSME_k}
	(\ref{item:Rxk_b})} of the CSME algorithm (cf.\ Section \ref{sec:CSME}).
Large adaption steps are facilitated by large entries in $\mat{P}_{x_k}$; thus, the proposed threshold is logical.
For a nominal situation, the mismatch between actual map  $\vec{\mu}^\mathfrak{M}$ and revised map $\vec{\widetilde{\mu}}^\mathfrak{M}$ vanishes as additional information is collected, yielding a noisy, nearly zero-mean signal $\mu_{j_\infty}^\mathfrak{M} - \widetilde{\mu}_{j_\infty}^\mathfrak{M}$.
For the faulty situation, the signal characteristic completely changes, which might serve as an indication whether for manual monitoring or some augmented and automated fault detection schemes, the investigation of which is beyond the scope of this paper.

\subsection{Performance Map Monitoring}
\label{sec:mapMonitor}

\begin{figure*}
	\centering
	\includegraphics[height=0.93\textheight]{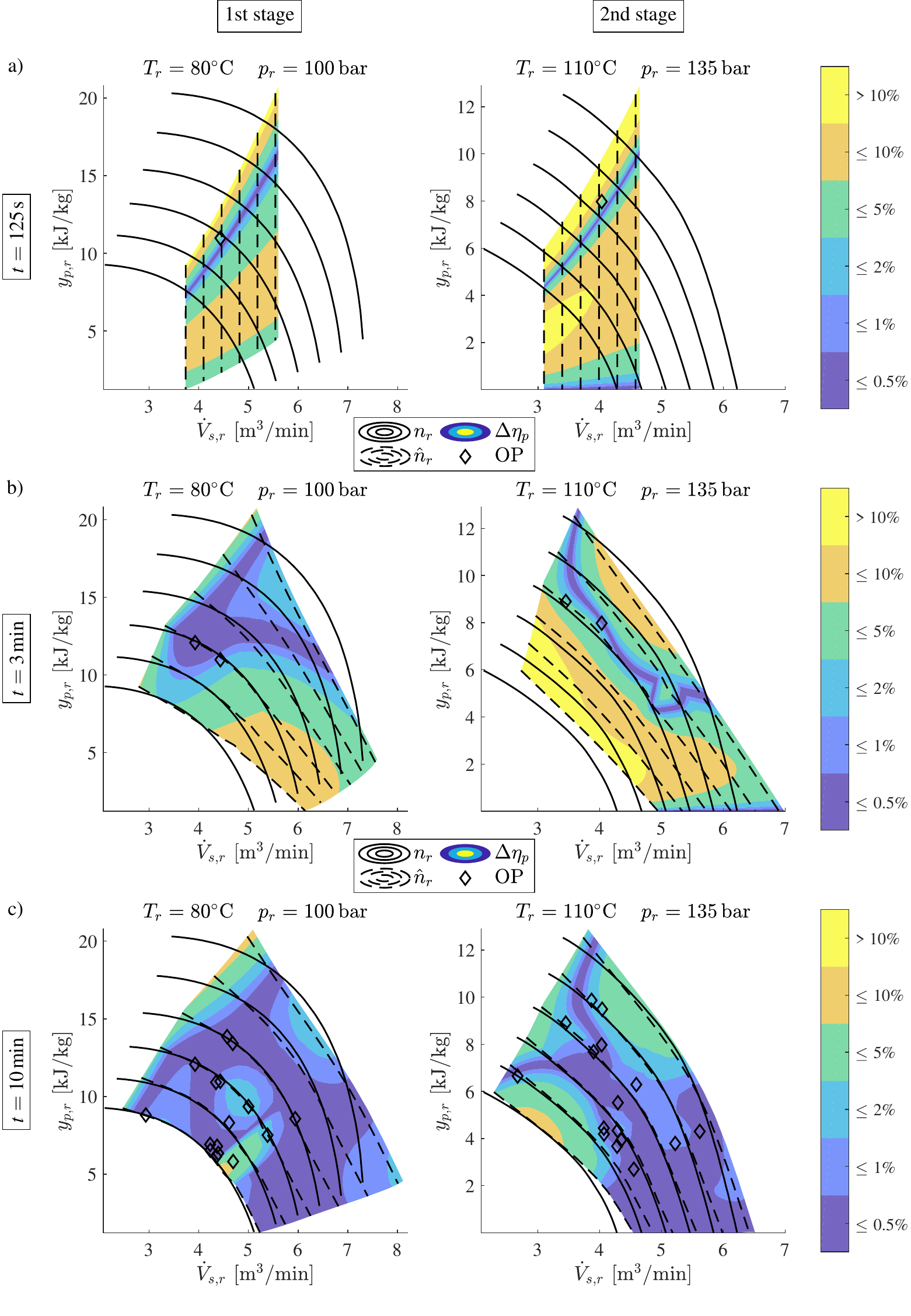}
	\caption[RME1]{True and estimated (revised) performance map of the first (left column) and second (right column) compressor stages in relation to the reference suction conditions ($T_r$, $p_r$) at three time stamps (rows); every \SI{30}{\second} (after enabling map estimation at $t=\SI{125}{\second}$), a new (estimated) operating point (OP) is drawn, shown by the increasing number of \raisebox{0pt}{\tikz{\draw[-,thick,scale=0.4](0em,0em)--(-0.7em,1em)--(0,2em)--(0.7em,1em)--(0em,0em);}} symbols; the filled contours, projected into the estimated performance map shape, represent levels of the efficiency estimation error $\Delta \eta_p = \left| \eta_p - \hat{\eta}_p \right|$ }
	\label{fig:RME1}
\end{figure*}

As denoted in Fig.\ \ref{fig:SE_FDI}, we utilize the delayed activation of the RME, which was suggested in step \mbox{%\ref{item:CSME_k}.
	(\ref{item:startRLS})} of Section \ref{sec:CSME}.
The preset ``threshold'' of \SI{2}{\percent} enables the map estimation after \SI{125}{\second}.
For this time stamp, the initialized map estimate (dashed speed curves), free of any reasonable \textit{a priori} shape, can be found in Fig.\ \ref{fig:RME1}a.
In this figure, polytropic work $y_p$ and suction volume flow $\dot{V}_s$ are displayed instead of their dimensionless counterparts.
Thus far, the information of merely one operating point (OP) marked by \raisebox{0pt}{\tikz{\draw[-,thick,scale=0.4](0em,0em)--(-0.7em,1em)--(0,2em)--(0.7em,1em)--(0em,0em);}} has been considered.
The introduced regularization terms provide the ``straight'' shape, since gradients ($\mat{L}_g$) and curvatures ($\mat{L}_c$) have a relative high cost to this moment.

\begin{figure*}
	\centering
	\includegraphics[height=0.93\textheight]{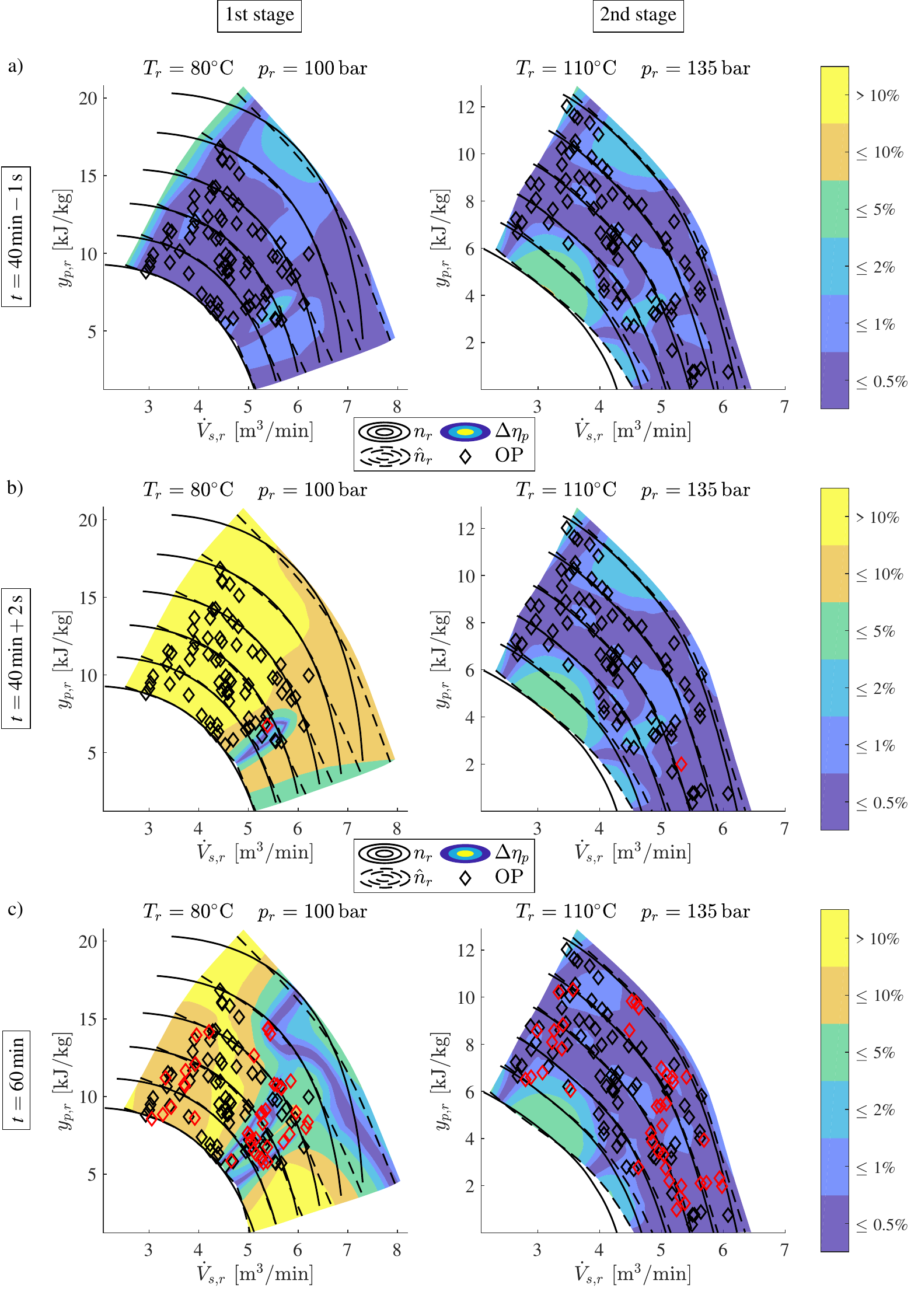}
	\caption[RME2]{Nomenclature, see above; additionally, operating points (OP) that occur after the fault event at $t=\SI{40}{\minute}$ are highlighted in red}
	\label{fig:RME2}
\end{figure*}

The presented shape is mapped into the typical \mbox{$y_p$--$\dot{V}_s$--$n$} diagram.
In contrast, the true map and the (actual and revised) map estimate\footnote
{
	For the presentation of estimates in Fig.\ \ref{fig:RME1} and Fig.\ \ref{fig:RME2}, the \textit{revised} map is used (cf.\ Section \ref{sec:CSME}).
}
are preserved in the dimensionless counterpart \mbox{$\Psi_p$--$\varphi$--$\Mu$}, which is not really suitable for the intended presentation here.\footnote
{
	The $\mu$-dimension becomes visible via $\eta_p$ or, to be precise, via $\Delta \eta_p$ (filled contours, an explanation follows).
}
Due to the basic model assumption that presupposes a static dependency between \mbox{$\Psi_p$--$\varphi$--$\mu$--$\Mu$} (cf.\ Section \ref{sec:stagemap}), it is necessary to use an artificial, time-invariant suction condition (index $r$) for conversion (\mbox{$\Psi_p$--$\varphi$--$\Mu$} $\rightarrow$ \mbox{$y_p$--$\dot{V}_s$--$n$})\footnote
{
	The real-gas models differ, whether the conversion refers to the true or the estimated map; see Section \ref{sec:setup}.
} according to relations (\ref{eq:Psi})--(\ref{eq:Mu2}) if the true performance map \mbox{$y_p$--$\dot{V}_s$--$n$} (solid speed curves) is to be fixed for the supervision monitor.
%To obtain a fixed (time-invariant) true performance map \mbox{$y_p$--$\dot{V}_s$--$n$} (solid speed curves), an artificial, time-invariant suction condition (index $r$) is used for conversion according to relations (\ref{eq:Psi})--(\ref{eq:Mu2}).\footnote
%{
%	The real-gas models differ, whether the conversion is concerned to the true or the estimated map, see Section \ref{sec:setup}.
%%}
As a result, the projected OP (related to $T_r$ and $p_r$) of the first and second stages do not lie on the same (projected) speed curve, even if the stages are mechanically coupled ($\Sa{n}=\Sb{n}$, cf.\ Fig.\ \ref{fig:SE_Input}a).
Note that both shapes, true and estimated, are depicted for the same $\Psi_p$ and $\Mu$ domain.

As the operating point varies, the shape adapts quickly, as can be seen in Fig.\ \ref{fig:RME1}b.
Here, \SI{55}{\second} after map initialization, 55 respective measurements have been incorporated.
Every \SI{30}{\second}, the corresponding OP estimated is drawn into the present diagram.
As a result of the flexible interpolation-based mapping scheme, it seems unreasonable to expect a correct shape adjustment for map regions in which no information has been collected thus far.
Consequently, the speed curve shape of the first stage does not converge correctly in the region where the surge line would be anticipated, as this region was not accessed thus far.
However, near the operating points, good estimation results can be seen.
Whenever a ``new'' region is entered, the supervisor may recognize this by the rapid decline of the LIL.

As time passes, and the amount of information collected increases (Fig.\ \ref{fig:RME1}c, Fig.\ \ref{fig:RME2}a), not only the estimated shape of speed curves but also the estimated shape of the efficiency map improves, which becomes obvious by the drawn levels of the efficiency estimation error $\Delta \eta_p = \left| \eta_p - \hat{\eta}_p \right|$, depicted as filled contours within the estimated speed curve shape.
Concerning these levels, a systematic estimation error remains, due to the different mapping functions used in the SE and the CSME, as mentioned in Section \ref{sec:setup}.
The fault event at $t=\SI{40}{\minute}$, a spontaneous decline in the first stage's $\mu$ map vectors of the reference (SE), affects neither the speed curve shapes nor the second stage's performance map at all (Fig.\ \ref{fig:RME2}b,c).
This complies with the correct behavior, facilitating the further improvement of the second stage's performance map beyond the fault event (cf.\ Fig.\ \ref{fig:RME2}c).
The low level of $\Delta \eta_p$ within the vicinity of the current (red) operating point, \SI{2}{\second} after the fault event (Fig.\ \ref{fig:RME2}b), proves the fast adaptability of the \textit{revised map}, which is presented for all time stamps in Fig.\ \ref{fig:RME1} and Fig.\ \ref{fig:RME2}.
A further issue, already stated, can be seen in Fig.\ \ref{fig:RME2}c for the first stage.
Within regions where much informations had already been collected (high concentration of past, black OP), the estimated map is highly inflexible.
However, if the fault had been detected, a reinitialization (after potential interventions) would be advisable anyway.

%%%%%%%%%%%%%%%%%%%%%%%%%%%%%%%%%%%%%%%%%%%%%%%%%%%%%%%%%%%%%%%%%%%%%%
\section{Conclusions}
\label{sec:conclusion}

From the methodological aspect, two main issues have been presented.

\paragraph*{Model building}
A novel, low-order dynamic model for centrifugal multi-stage compressors has been derived.
Real-gas behavior is taken into account explicitly.
To this end, the generic LKP real-gas equation of state \cite{lee1975LKP} is applied.
Several refinements are provided to embed this equation and its derivatives properly into the overall model scheme, in terms of accuracy and computational speed.
A compressor stage's behavior arises from its possibly time-variant performance map.
The proposed approach utilizes an interpolation scheme based on four grid vectors, i.e., polytropic head coefficient, flow coefficient, work input factor, and machine Mach number, which allows for the description of nearly arbitrary performance map shapes.
The proposed scheme may easily be extended by further dependencies, e.g., for variable inlet guide vanes.

\paragraph*{Monitoring}
The \textit{Unscented Kalman Filter} approach and a new \textit{Recursive Map Estimation} are combined, yielding a novel real-time estimation scheme, which is expected to be superior if the parameters to be estimated have a distinct operating point dependency, as is the case for the grid vectors of a compressor stage's performance map.
Real-time capability is addressed via 
\begin{itemize}[label=$\bullet$, leftmargin=*]
	\item 	a first-principle, but---in detail---approximate and consequently less computationally intensive model;
	\item 	a recursive formulation of all estimation steps, yielding a constant calculation workload;
	\item 	an optimal preservation of past estimates concerning the operating point dependency within fixed-size grid vectors, yielding a constant memory requirement.
\end{itemize}
As a by-product, three time-variant supplementary observations can be provided for the monitoring task in the context of monitoring of a multi-stage compressor:
\begin{enumerate}[leftmargin=*]
	\item 	a performance map for every compressor stage, i.e., the estimated shape of speed and efficiency curves;
	\item 	a \textit{Local Information Level}, indicating the reliability of estimates at the local operating point;
	\item 	a fault indicator for every compressor stage, which might be extended for fault detection and isolation if conceivable faults have been defined.
\end{enumerate}
The estimator is able to handle \textit{a priori} knowledge optionally, whether the task is to monitor deviations from the \textit{a priori} presumed performance map or to identify the performance map during operation.
\\
\\
The model-based monitoring scheme was validated via numerical simulations of a two-stage carbon dioxide compressor operating in the supercritical phase of the fluid.
The reference simulation, which replaces the real ``measurements'', was modified considerably; e.g., the real-gas model was interchanged, and the mass transportation delay was considered.
In spite of this adverse situation, the proposed estimator performed well.
The estimator was capable of tracking every state or variable, whether it was measured or not, without noticeable bias.
For operating ranges that have already been reached, the estimated performance maps converged correctly.
Within the remaining regions, the map shape maintains its flexibility.
A preset fault event was isolated (to the respective compressor stage) correctly, and the overall behavior of the estimates and fault indicators was as desired.

Subjects of future research may arise from the following:
\begin{itemize}[label=$\bullet$, leftmargin=*]
	\item 	The request to continuously incorporate an alteration, which might be detectable with the proposed scheme already, into a nearly converged map estimate.
			Since the integration of common (global) forgetting factors is considered unreasonable within the given context, the approach of local forgetting might be further investigated.
	\item 	The demand to recover from an erroneous map status, which may be triggered from faulty measurements or extreme deviations between plant and model behavior, e.g., due to a short period of compressor surge.
			Strategies for recovering as well as surge modeling, may contribute to this issue.
	\item 	Augmented fault detection and isolation schemes, i.e., the real-time classification of conceivable failure sets.
\end{itemize}

% Ausblick: Vergessensfaktoren...
% Ausblick: Recovery...
% Ausblick: Machine Learning FDI...

%%%%%%%%%%%%%%%%%%%%%%%%%%%%%%%%%%%%%%%%%%%%%%%%%%%%%%%%%%%%%%%%%%%%%%
\section*{Acknowledgment}
This work was supported by MAN Energy Solutions SE and the Federal Ministry for Economic Affairs and Energy based on a decision by the German Bundestag as part of the ECOFLEX-Turbo project [grant number 03ET7091T].

%%%%%%%%%%%%%%%%%%%%%%%%%%%%%%%%%%%%%%%%%%%%%%%%%%%%%%%%%%%%%%%%%%%%%%
% The bibliography is stored in an external database file
% in the BibTeX format (file_name.bib).  The bibliography is
% created by the following command and it will appear in this
% position in the document. You may, of course, create your
% own bibliography by using thebibliography environment as in
%
% \begin{thebibliography}{12}
% ...
% \bibitem{itemreference} D. E. Knudsen.
% {\em 1966 World Bnus Almanac.}
% {Permafrost Press, Novosibirsk.}
% ...
% \end{thebibliography}

% Here's where you specify the bibliography style file.
% The full file name for the bibliography style file 
% used for an ASME paper is asmems4.bst.

% Here's where you specify the bibliography database file.
% The full file name of the bibliography database for this
% article is asme2e.bib. The name for your database is up
% to you.
\bibliography{Gentsch_King_MapEstimation_bib}

%%%%%%%%%%%%%%%%%%%%%%%%%%%%%%%%%%%%%%%%%%%%%%%%%%%%%%%%%%%%%%%%%%%%%%
%\appendix       %%% starting appendix
%\section*{Appendix A:}
%\label{sec:appendixA}

\end{document}